\documentclass[fleqn,usenatbib]{mnras}

\usepackage{newtxtext}
\usepackage[varvw]{newtxmath}

\usepackage[T1]{fontenc}

\DeclareRobustCommand{\VAN}[3]{#2}
\let\VANthebibliography\thebibliography
\def\thebibliography{\DeclareRobustCommand{\VAN}[3]{##3}\VANthebibliography}

\usepackage{graphicx}	
\usepackage{amsmath}	
\usepackage{color}
\usepackage[flushleft]{threeparttable}

\newcommand{\phn}{\phantom{0}}
\newcommand{\phnn}{\phantom{00}}

\newcommand{\phs}{\phantom{$>$}}

\usepackage{enumitem}
\newlist{findings}{enumerate}{1}
\setlist[findings,1]{
  label={--},
  leftmargin=*,
  align=left,
  labelsep=2mm,
}

\title[Disc-jet system around W75N(B)-VLA3]{A compact Keplerian-like disc and H30$\alpha$ emission towards the radio jet in the massive protostar W75N(B)-VLA3}

\author[\'A. S\'anchez-Monge et al.]
{\'A.\ S\'anchez-Monge,$^{1,2}$\thanks{E-mail: asanchez@ice.csic.es (\'A.~S\'anchez-Monge)}
J.~F.\ G\'omez,$^{3}$
J.~M.\ Torrelles,$^{1,2}$
S.\ Curiel,$^{4}$
J.~M.\ Girart,$^{1,2}$
G.\ Surcis,$^{5}$
\newauthor
C.\ Carrasco-Gonz\'alez,$^{6}$
G.\ Anglada,$^{3}$
G.~A.\ Fuller,$^{7,8}$
C.\ Goddi,$^{5,9,10}$
W.~H.~T.\ Vlemmings,$^{11}$
\newauthor
A.~R.\ Rodr\'iguez-Kamenetzky,$^{12}$
H.~J.\ van Langevelde,$^{13,14}$
J.-S.\ Kim,$^{15,16}$ 
S.-W.\ Kim,$^{16}$
and J.\ Cant\'o$^{4}$
\\
\\ 
$^{1}$Institut de Ci\`encies de l'Espai (ICE), CSIC, Campus UAB, Carrer de Can Magrans s/n, E-08193, Bellaterra, Barcelona, Spain\\
$^{2}$Institut d'Estudis Espacials de Catalunya (IEEC), E-08860, Castelldefels, Barcelona, Spain\\
$^{3}$Instituto de Astrof{\'i}sica de Andaluc{\'i}a (IAA-CSIC), Glorieta de la Astronom{\'i}a s/n, E-18008 Granada, Spain\\
$^{4}$Instituto de Astronom\'{i}a, Universidad Nacional Aut\'onoma de M\'{e}xico (UNAM), Apdo Postal 70-264, Ciudad de M\'{e}xico, Mexico\\
$^{5}$INAF - Osservatorio Astronomico di Cagliari, Via della Scienza 5, I-09047, Selargius, Italy\\
$^{6}$Instituto de Radioastronom\'ia y Astrof\'isica (IRyA-UNAM), Morelia, Mexico\\
$^{7}$Jodrell Bank Centre for Astrophysics, Department of Physics \& Astronomy, The University of Manchester, Oxford Road, Manchester M13 9PL, UK\\
$^{8}$I.\ Physikalisches Institut, Universit\"{a}t zu K\"{o}ln, Z\"{u}lpicher Str.\ 77, D-50937 K\"{o}ln, Germany\\
$^{9}$Instituto de Astronomia, Geof\'isica e Ci\^encias Atmosf\'ericas, Universidade de S\~ao Paulo, R. do Mat\~ao, 1226, S\~ao Paulo, SP 05508-090, Brazil\\
$^{10}$Dipartimento di Fisica, Universit\'a degli Studi di Cagliari, SP Monserrato-Sestu km 0.7, I-09042 Monserrato (CA), Italy\\
$^{11}$Department of Space, Earth and Environment, Chalmers University of Technology, 412 96, Gothenburg, Sweden\\
$^{12}$Instituto de Astronom\'{\i}a Te\'orica y Experimental (IATE, CONICET-UNC), C\'ordoba, Argentina\\
$^{13}$Joint Institute for VLBI ERIC, Oude Hoogeveensedijk 4, 7991 PD Dwingeloo, The Netherlands\\
$^{14}$Sterrewacht Leiden, Leiden University, Postbus 9513, 2300 RA Leiden, The Netherlands\\
$^{15}$National Astronomical Observatories, Chinese Academy of Sciences, 20A Datun Road, Chaoyang District, Beijing, China\\
$^{16}$Korea Astronomy and Space Science Institute, 776 Daedeok-daero, Yuseong-gu, Daejeon 34055, Republic of Korea
}

\date{Accepted XXX. Received YYY; in original form ZZZ}

\pubyear{2025}

\begin{document}
\label{firstpage}
\pagerange{\pageref{firstpage}--\pageref{lastpage}}
\maketitle

\begin{abstract}
Studying young, high-mass stellar objects is challenging for testing models of massive star formation due to their great distances, often kiloparsecs away. This requires extremely high-angular resolution to resolve features like accretion discs around massive protostars. Consequently, while powerful, collimated outflows are evident in massive protostars, the compact accretion discs anticipated at their centres are still proving difficult to pinpoint. This study presents ALMA continuum and molecular line observations at 1.3~mm of the massive protostar W75N(B)-VLA3. The observations achieve an angular resolution of $\sim$$0\farcs12$ ($\sim$156~au). Dense gas tracers reveal a circumstellar disc of $\sim$450~au in radius surrounding VLA3, with an orientation perpendicular to its associated thermal radio jet. From the millimetre continuum, a total mass of $\approx$0.43--1.74~$M_\odot$ is estimated for the disc. The disc's velocity profile is consistent with Keplerian rotation around a protostar of $\approx$16~$M_\odot$. This adds VLA3 to the small number of massive disc-protostar-jet systems documented in the literature with a centrifugally supported disc with a radius less than 500~au. Additionally, we detected H30$\alpha$ recombination line emission towards the radio jet powered by VLA3. Despite limitations in the spatial and spectral resolution, our data reveal a very broad line, indicative of high-velocity motions, as well as a tentative velocity gradient in the jet's direction, thus favouring the H30$\alpha$ emission to originate from the radio jet. Should this interpretation be confirmed with new observations, W75N(B)-VLA3 could represent the first protostellar radio jet for which a thermal radio recombination line has been detected.
\end{abstract}




%
\section{Introduction}\label{sec:introduction}

The high-mass star-forming region (HMSFR) W75N(B) is a well-studied region that contains three massive protostars detected at radio continuum and millimetre wavelengths: VLA1, VLA2, and VLA3 \citep[e.g.,][]{Hunter1994, Torrelles1997, Gomez2023}. These protostars, with luminosities equivalent to B1-B0.5 type stars \citep[e.g.,][]{Torrelles1997, Shepherd2001, Anglada2018}, are located within a region of $\sim$1.5 arcsec \citep[$\sim$1950~au at the distance of 1.3~kpc, as estimated by][]{Rygl2012}, where two filaments seen in NH$_3$ seem to be interacting \citep[][]{CarrascoGonzalez2010a}. Previous studies have reported outflow activity observed in radio continuum, maser emission (predominantly associated with VLA1 and VLA2), and thermal molecular lines \citep[e.g.,][]{Torrelles1997, Torrelles2003, Kim2013, Surcis2014, Surcis2023, CarrascoGonzalez2015, Colom2018, Colom2021, RodriguezKamenetzky2020, Gomez2023}. 

These studies have revealed that the three massive protostars, at distinct evolutionary stages, exhibit a variety of characteristics. Specifically, VLA1 drives a thermal radio jet at scales of $\sim$0.1~arcsec ($\sim$130~au) and shows signs of early stage photoioinisation. The outflow associated with VLA2, observed in both water (H$_2$O) maser emission and radio continuum, has undergone a remarkable transformation over several decades. Initially, it was almost isotropic, but has since become collimated at scales of $\sim$0.2~arcsec \citep[$\sim$260~au; see][for a review of the properties of VLA1 and VLA2]{Surcis2023}. Completing the picture, VLA3 drives a compact thermal radio jet at scales of a few hundred au \citep[see][]{CarrascoGonzalez2010a}. This protostar is thought to be responsible for exciting the two pairs of obscured Herbig-Haro (HH) objects (Bc[E]-Bc[W], Bd-VLA4) detected at radio wavelengths $\sim$4~arcsec ($\sim$5200~au) south of VLA3. These HH objects exhibit proper motions moving outwards from the protostar at velocities exceeding 100~km~s$^{-1}$ \citep[see][for a detailed description of all these properties]{RodriguezKamenetzky2020, Gomez2023}. Relatively weak H$_2$O maser emission has also been reported towards VLA3 \citep[60~mJy, V$_{\rm LSR}$ = 10.5~km~s$^{-1}$;][]{Torrelles1997}. Since VLA1, VLA2, and VLA3 share a common origin from the gas and dust environment of W75N(B), this region has become a prime candidate for testing high-mass star formation models.

While different theoretical scenarios have been proposed to explain the formation of high-mass stars (e.g., competitive accretion in a turbulence-dominated core, \citealt{Krumholz2009}; competitive accretion driven by a stellar cluster, \citealt{BonnellBate2006}; Bondi-Hoyle accretion, \citealt{Keto2007}), all of them predict the formation of circumstellar discs that enable the transport of mass from the envelope/environment to the forming star \citep[see, e.g.,][for different reviews on high-mass star formation]{Tan2014, Krumholz2015, Motte2018, Zhao2020, Avison2023, Beuther2025}. Therefore, at small scales, one expects to find massive protostars surrounded by discs, accompanied by collimated jets, similar to those observed in low-mass star formation. In their comprehensive review, \citet{BeltrandeWit2016} summarise the evidence and properties for discs around intermediate-mass ($\sim$2--7~$M_{\odot}$) and high-mass ($\gtrsim$7~$M_{\odot}$) stars \citep[see also][]{Zhao2020, Ahmadi2023}. Based on this study, while there is little doubt that discs exist around intermediate-mass stars, discs around the most massive stars still remain elusive. Indeed, even though powerful collimated outflows are identified toward several massive protostars, the expected compact accretion discs at their centres are still hard to find \citep[e.g.,][]{Goddi2020}. Observations tend to identify massive, large, rotating structures, usually referred to as `toroids' with sizes $\gtrsim$5000~au \citep[e.g.,][]{Torrelles1983, Beltran2005, Beltran2011}. A few notable examples of massive protostars associated with compact disc--outflow systems, recently discovered thanks to the improved resolving power of current interferometers, include AFGL\,4176 \citep[][]{Johnston2015, Johnston2020}, Cepheus~A-HW2 \citep[][]{Patel2005, Curiel2006, JimenezSerra2007, Sanna2025}, G11.92$-$0.61\,MM1 \citep[][]{Ilee2018, Bayandina2025}, G17.64+0.16 \cite[][]{Maud2018, Maud2019}, G23.01$-$0.41 \citep[][]{Sanna2019}, G35.20$-$0.74\,N \citep[][]{SanchezMonge2013a, SanchezMonge2014, Beltran2016}, GGD\,27-MM1 \citep[][]{CarrascoGonzalez2012, Girart2018, FernandezLopez2023}, IRAS~16547$-$4247 \citep[][]{Rodriguez2005, Zapata2019}, IRAS~20126$+$4104 \citep[][]{Cesaroni1999, Cesaroni2025, Palau2017}, Orion~Src\,I \citep[][]{Hirota2017, Ginsburg2018}, and W75N(B)-VLA2 \citep[][]{Torrelles1997, CarrascoGonzalez2015, Gomez2023}. For other cases, see \citet{Ginsburg2023} and references therein.

In the recent paper by \cite{Gomez2023}, we presented ALMA observations of the continuum, SiO and H$_2$CO emission lines at 1.3~mm wavelengths towards the HMSFR W75N(B). These observations detected the VLA1, VLA2, and VLA3 protostars at millimetre continuum wavelengths at a superb angular resolution and sensitivity. The H$_2$CO emission exhibits a fragmented structure surrounding the three massive protostars, while the SiO emission is strongly concentrated towards VLA2. The elongated structure of the SiO emission, perpendicular to the VLA2 thermal radio jet, together with  its kinematics, led the authors to interpret it as arising in strong shocks within a toroid and a wide-angle outflow encircling this protostar.

In the current work, we now shift our focus to VLA3, which is the brightest high-mass protostar at millimetre wavelengths in W75N(B). We undertake a detailed analysis of the spatio-kinematical distribution of different molecular dense gas tracers, together with shocks and ionised gas tracers, and the mm-continuum around VLA3. In Sect.~\ref{sec:observations}, we provide a brief summary of the ALMA observations carried out by \cite{Gomez2023}. The main results obtained towards VLA3 are presented in Sect.~\ref{sec:results}, while Sect.~\ref{sec:discussion} discusses these findings. Finally, Sect.~\ref{sec:conclusions} summarises our main conclusions.

\begin{table}
\centering
\caption{\label{tab:spectral-setup}Spectral setup of the ALMA observations}
\begin{tabular}{l c c c}
\hline\hline \noalign{\smallskip} 
  Spectral window --- 
& Central Freq.
& Bandwidth
& Channel width 
\\
  main targeted species
& (MHz)
& (MHz)
& (km~s$^{-1}$)
\\
\hline \noalign{\smallskip} 
spw~0 --- continuum & 231805.970 & 1875.0\phn & 20.0\phnn    \\
spw~1 --- $^{12}$CO & 230538.000 & \phn468.75 & \phn0.734 \\
spw~2 --- $^{13}$CO & 220398.684 & \phn234.38 & \phn0.384 \\
spw~3 --- C$^{18}$O & 219560.358 & \phn234.38 & \phn0.385 \\
spw~4 --- H$_2$CO   & 218222.192 & \phn468.75 & \phn0.775 \\
spw~5 --- SiO       & 217104.980 & \phn468.75 & \phn0.779 \\
\hline
\end{tabular}
\end{table}

\begin{figure*}
\centering
\includegraphics[width=1.0\textwidth]{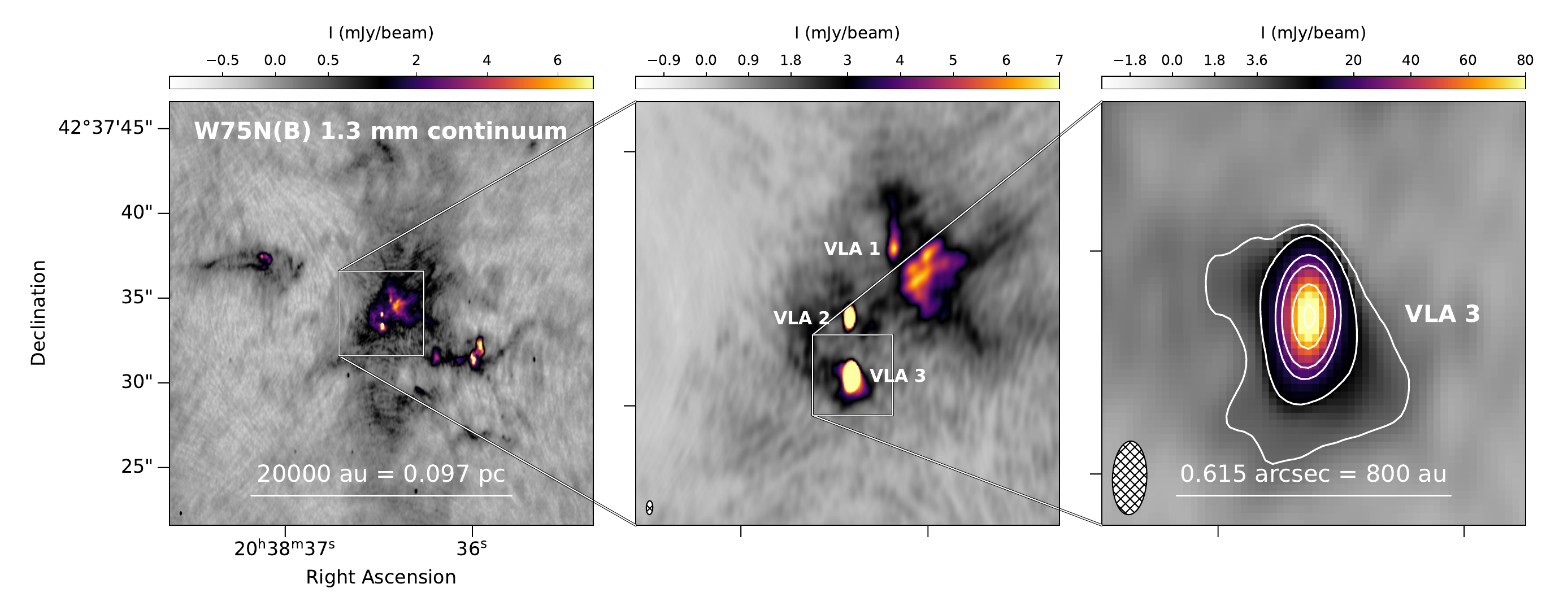}
\vspace{-0.5cm}
\caption{ALMA 1.3~mm continuum emission image of the W75N(B) high-mass star-forming complex \citep[see also][]{Gomez2023}. From left to right, the panels show the whole region; the inner $5\arcsec\times5\arcsec$ ($\simeq0.03~{\rm pc}\times0.03~{\rm pc}$) area engulfing the three main massive protostars (VLA1, VLA2 and VLA3); and a close-up view of the $0\farcs95\times0\farcs95$ ($\simeq1230~{\rm au}\times1230~{\rm au}$) centered at VLA3. The white contours in the right panel show the ALMA 1.3~mm continuum emission at 3, 6, 18, 30, 60, and 90~mJy~beam$^{-1}$ (beam $=0\farcs165\times0\farcs077$, PA$=-2\fdg2$; $\mathrm{rms}\simeq0.08$~mJy~beam$^{-1}$), and are used to emphasize the extended emission that exists around the compact core of VLA3. The hatched ellipse in the bottom-left corner of the panels depicts the synthesised beam.}
\label{fig:continuum}
\end{figure*}

%
\section{Observations}\label{sec:observations}

We used ALMA to observe the continuum emission at 1.3~mm and multiple spectral lines towards W75N(B)-VLA3. The observations were carried out on April 29, August 18, and October 28, 2021, and on August 2, 2022, as part of project 2019.1.00059.S. A full description of the setup of the ALMA observations as well as the calibration and reduction procedures is given in \cite{Gomez2023}. The phase centre of the observations is $\alpha$(J2000) = 20$^\mathrm{h}$38$^\mathrm{m}$36$.\!^\mathrm{s}$486, $\delta$(J2000) = 42$^{\circ}$37$^\prime$34$.\!\!^{\prime\prime}$09. The spectral setup (see Table~\ref{tab:spectral-setup}) consists of a broadband (1.875~GHz wide) spectral window (spw\,0) with low spectral resolution, intended to detect the continuum emission, and five additional high-spectral resolution windows, centred at selected spectral line transitions, and aimed at studying the gas kinematics. The total effective bandwidth ($\sim$3.75~GHz) includes additional spectral lines (see Sect.~\ref{sec:molecules}).

For imaging, the task \texttt{tclean} of CASA \citep[version 6.2.1.7; ][]{CASA2022} was used with a Brigg's weighting of visibilities and robust parameter of 0.5, obtaining a synthesized beam size = $0\farcs165\times0\farcs077$, with position angle (PA) $=-2\fdg2$, and an rms $\simeq$ 0.08~mJy~beam$^{-1}$ for the continuum image. For the spectral line data the rms is $\simeq$ 1~mJy~beam$^{-1}$ per velocity channel of 0.78~km~s$^{-1}$, with synthesised beams comparable to that of the continuum image. All images presented here have been obtained after self-calibration and corrected by the primary beam response, which has a full-width at half-maximum (FWHM) $\simeq$ 25$^{\prime\prime}$ at 230~GHz \citep[see][for more details]{Gomez2023}.

%
\section{Results}\label{sec:results}

\begin{figure*}
\centering
\includegraphics[width=1.0\textwidth]{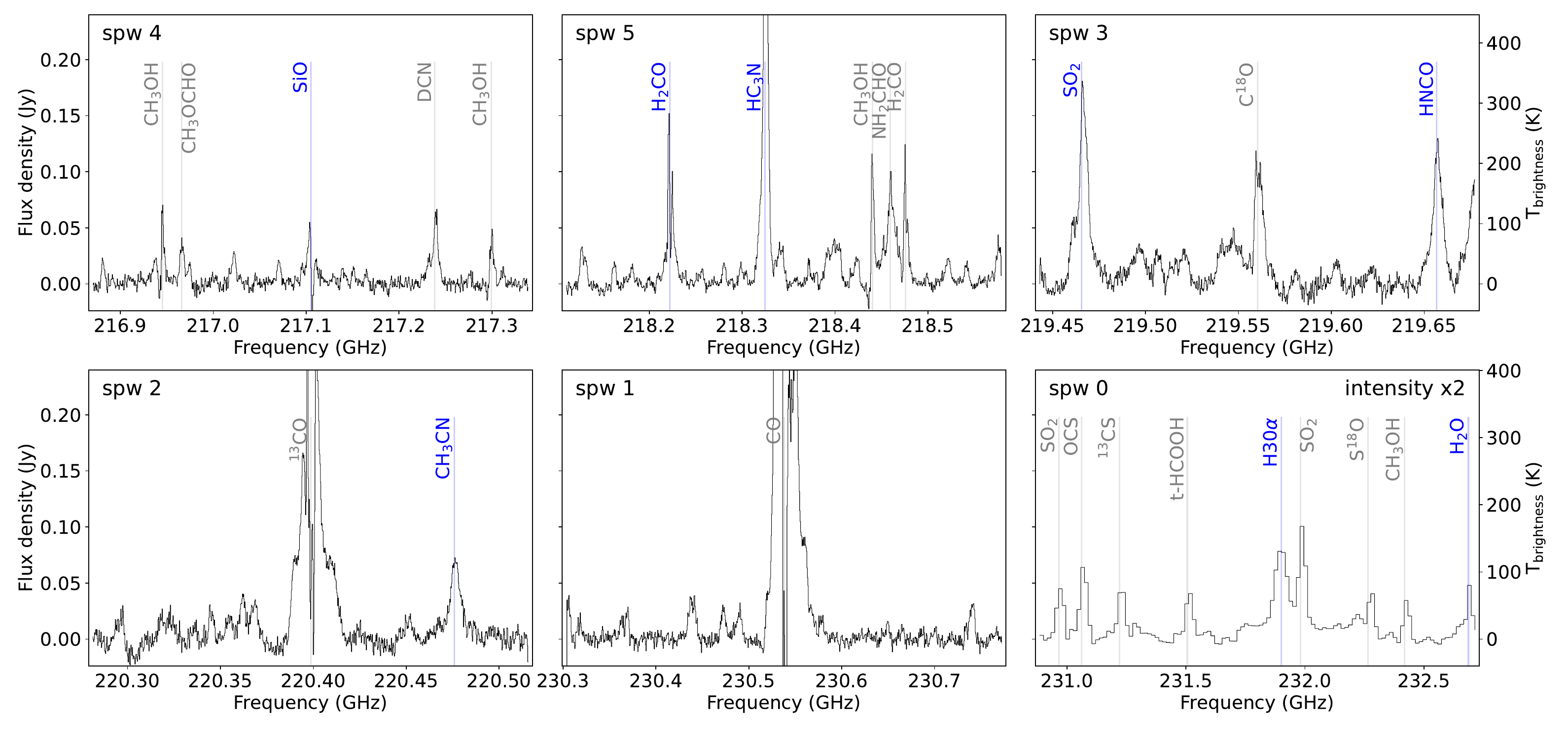}
\vspace{-0.5cm}
\caption{ALMA spectra extracted towards the young stellar object VLA3, located at $\alpha(\mathrm{J2000})=20^\mathrm{h}38^\mathrm{m}36.4815^\mathrm{s}$ and $\delta(\mathrm{J2000})=+42^{\circ}37^{\prime}33.355^{\prime\prime}$, after integrating over the region defined by the 3~mJy~beam$^{-1}$ contour level of the 1.3~mm continuum emission (see right panel of Fig.~\ref{fig:continuum}). The six panels show the different spectral windows observed with ALMA from low to high frequency \citep[see the spectral setup in][and Table~\ref{tab:spectral-setup}]{Gomez2023}. The bottom-right panel shows the low spectral resolution data from spectral window `spw\,0', whose emission has been multiplied by a factor of 2. We have identified some of the most prominent detected spectral lines with the corresponding species. The spectral lines highlighted in blue are discussed in this work. A complete molecular inventory of the W75N(B) region obtained with our ALMA data will be discussed in a forthcoming paper (S\'anchez-Monge et al., in preparation).}
\label{fig:spectra}
\end{figure*}

\begin{table*}
\centering
\caption{\label{tab:spectral-lines}Selection of bright spectral lines detected towards VLA3 (see blue markers in Fig.~\ref{fig:spectra}), and studied in this work}
\begin{tabular}{l l c c c c l}
\hline\hline 
  Species
& transition
& Freq.~(MHz)
& $\log_\mathrm{10}$~[$A_{\mathrm{ij}}$~(s$^{-1}$)]
& $E_\mathrm{up}$~(K)
& $n_\mathrm{crit}$~(cm$^{-3}$)
& tracer of \ldots
\\
\hline 
SiO      & $5$--$4$                           & 217104.9190 & $-3.28429$ & \phnn31.26 & \phs$2.5\times10^{6}$ & shocked gas \\
H$_2$CO  & $3_{0,3}$--$2_{0,2}$               & 218222.1920 & $-3.55007$ & \phnn20.96 & \phs$3.0\times10^{6}$ & dense gas \\
HC$_3$N  & $24$--$23$                         & 218324.7230 & $-3.08296$ & \phn130.98 & \phs$1.7\times10^{7}$ & dense gas \\
SO$_2$   & $22_{2,20}$--$22_{1,21}$ ; v$_2$=1 & 219465.5447 & $-3.99721$ & 1012.74    & \ldots & hot, dense gas\\
HNCO     & $10_{3,8}$--$9_{3,7}$              & 219656.7695 & $-3.92039$ & \phn432.96 & \ldots & dense gas\\
         & $10_{3,7}$--$9_{3,6}$              & 219656.7708 & $-3.92039$ & \phn432.96 & \ldots & dense gas \\
CH$_3$CN & $12_{8}$--$11_{8}$                 & 220475.8078 & $-3.29104$ & \phn525.57 & $>$$1.9\times10^{7}$ & dense gas \\
H        & $31$--$30$ (30$\alpha$)            & 231900.9280 & \ldots     & \ldots     & \ldots & ionised gas \\
H$_2$O   & $5_{5,0}$--$6_{4,3}$ ; v$_2$=1     & 232686.7000 & $-5.32146$ & 3461.91    & \ldots & hot, dense gas \\
\hline
\end{tabular}
\begin{tablenotes}
\footnotesize
\item {\bf Notes}. The Einstein $A_{\mathrm{ij}}$ coefficients
and the upper energy levels $E_\mathrm{up}$ are obtained from the Cologne Database for Molecular Spectroscopy \citep[CDMS;][]{Mueller2001, Mueller2005} which is also available through the Virtual Atomic and Molecular Data Centre \citep[VAMDC;][]{Endres2016}. The critical densities $n_\mathrm{crit}$ are derived using the collisional rates from the Leiden Atomic and Molecular Database \citep[LAMDA;][]{Schoeier2005} for those transitions with available information.
\end{tablenotes}
\end{table*}

%
\subsection{Dust 1.3~mm-continuum emission}\label{sec:continuum}

Continuum images at a wavelength of 1.3~mm of the HMSFR W75N(B) are shown in Fig.~\ref{fig:continuum} through zoomed-in views from the whole region of $25\arcsec\times25\arcsec$ ($\simeq0.16~{\rm pc}\times0.16~{\rm pc}$), to a region of $5\arcsec\times5\arcsec$ ($\simeq0.03~{\rm pc}\times0.03~{\rm pc}$) centered on VLA2, and down to a region of $0\farcs95\times0\farcs95$ ($\simeq1230~{\rm au}\times1230~{\rm au}$) around VLA3, which is the brightest millimetre continuum source in W75N(B) \citep[see][]{Gomez2023}. VLA3 shows very compact, bright continuum emission, with an observed  size at FWHM ($0\farcs18\times0\farcs09$; PA $=-2^{\circ}$) comparable to the synthesized beam size ($0\farcs17\times0\farcs08$; PA $=-2^{\circ}$). The maximum intensity of the continuum emission in VLA3 is $I_\nu = 97$~mJy~beam$^{-1}$; slightly different from that reported by \cite{Gomez2023} measured with a smaller beam ($I_\nu$ = 87~mJy~beam$^{-1}$; beam $= 0\farcs13\times0\farcs06$, PA $=-2^{\circ}$), obtained by excluding baselines shorter than 200~k$\lambda$ to better identify weak compact sources from extended emission. There is also relatively weak extended emission surrounding VLA3 at a level of $\sim$3~mJy~beam$^{-1}$ (corresponding to about 30$\sigma$), and covering a region of $\sim$$0\farcs5$, extending preferentially in the northeast-southwest direction. The total flux density of VLA3 is $S_\nu\simeq149\pm1$~mJy, measured within the 3~mJy~beam$^{-1}$ contour level shown in Fig.~\ref{fig:continuum}-right, which corresponds to an area of $\sim$0.16~arcsec$^{2}$ and an effective circular radius of $\sim$$0\farcs23$ ($\sim$300~au).

Based on the flux density measured at a wavelength of 3.6~cm for the free-free emission of the thermal radio jet associated with VLA3 (4~mJy), and assuming a spectral index $\alpha=+0.6$ ($S_\nu\propto\nu^\alpha$) for the free-free continuum emission \citep{CarrascoGonzalez2010a}, the expected contribution of this component can be extrapolated to 1.3~mm, yielding a value of $\sim$30~mJy. Subtracting this contribution from the total emission observed at 1.3~mm infers a residual flux density of $\sim$120~mJy, attributable to thermal dust emission.

Assuming that the dust continuum emission at 1.3~mm is optically thin and the dust is isothermal within the observed region around VLA3, the dust mass content can be estimated from the expression:
\begin{equation}
{M_{\rm dust} = \frac{S_\nu d^2}{\kappa_\nu B_\nu(T_{\rm dust})}},
\end{equation}
where $d$ is the distance to the source, $\kappa_\nu$ the dust opacity, and $B_\nu(T_{\rm dust})$ the Planck function at the dust temperature ($T_{\rm dust}$). Then, adopting as in \cite{Gomez2023} a dust opacity  $\kappa_\nu$(1.3~mm) = 0.899~cm$^2$~g$^{-1}$ \citep[see][]{OssenkopfHenning1994}, $d = 1.3$~kpc, $T_{\rm dust}=50$~K, and gas-to-dust ratio of 100, the following expression for the total mass of dust+gas is obtained:
\begin{equation} \label{eq:massdust}
{M_{\rm dust+gas} =  1.4\times 10^{-2}~{M}_{\odot} \left[\frac{S_{\nu} }{\rm mJy}\right] \times  \left[\frac{T_{\rm dust}}{50~{\rm K}}\right]^{-1}}.
\end{equation}

We derive $M_{\rm dust+gas}$ $\approx$ 1.74~$M_{\odot}$ from the measured flux density of 120~mJy for thermal dust emission, assuming $T_{\rm dust}=$ 50~K. This mass reduces to $\approx$ 0.43~$M_\odot$ if a higher dust temperature of 200~K is considered. Such a high temperature is expected in the inner regions of the source, based on the detection of vibrationally excited lines of different molecular species (see Sect.~\ref{sec:molecules}). This temperature range aligns well with the expected values derived from relations established by \citet{Tobin2020} for discs associated with low-mass protostars, following extrapolation to VLA3's properties. We note that if the dust emission is not optically thin, the value of $M_{\rm dust+gas}$ will likely be higher \citep[see e.g.,][]{AnezLopez2020}. Despite this, the inner, optically-thick portion of the core would also be expected to be hotter than the temperatures we considered, which would yield lower masses and partially compensate for the opacity effects. Given the current spatial resolution and the fact that detailed modelling of the density and temperature structure of the dust surrounding VLA3 is beyond the scope of this work, we have used the plausible range of temperatures indicated above.

From its continuum emission of $\approx4$~mJy at 3.6~cm \citep[see][]{CarrascoGonzalez2010a, RodriguezKamenetzky2020} and the empirical correlation between bolometric luminosity and radio continuum luminosity at centimetre wavelengths \citep[see][and their Eq.~28]{Anglada2018}, a luminosity of $\sim4\times10^4$~$L_{\odot}$ is estimated for VLA3, equivalent to a stellar mass of 17--19~$M_{\odot}$ \citep[from the mass-luminosity relation for ZAMS stars, e.g.][]{Salaris2005, Graefener2011}. This stellar mass is about one order of magnitude larger than the dust+gas mass derived for the dense core, but comparable to the dynamical stellar mass of $\sim$16~M$_{\odot}$ (see Sect.~\ref{sec:pvplots}).

%
\subsection{Molecular emission}\label{sec:molecules}

Figure~\ref{fig:spectra} presents ALMA spectra extracted towards VLA3. The flux density of these spectra was obtained by integrating over the region defined by the 3~mJy~beam$^{-1}$ contour level of the 1.3~mm continuum emission (see Fig.~\ref{fig:continuum}-right panel). The figure comprises six panels, each displaying a different spectral window observed with ALMA (see Table~\ref{tab:spectral-setup}), arranged in order of increasing frequency. Some of the most prominent spectral lines detected within these windows have been identified, and assigned to their corresponding molecular species. The spectral lines highlighted in blue in this figure are those specifically discussed below, in the context of the study on VLA3, and listed in Table~\ref{tab:spectral-lines}.

A comprehensive analysis of the complete molecular line identification and inventory in the W75N(B) HMSFR will be presented in a forthcoming publication (Sánchez-Monge et al., in preparation). This study will focus on a region of $\sim$30 arcseconds ($\sim$39,000~au), encompassing the 40 millimetre continuum sources previously reported by \cite{Gomez2023}.

\begin{figure*}
\centering
\includegraphics[width=1.0\textwidth]{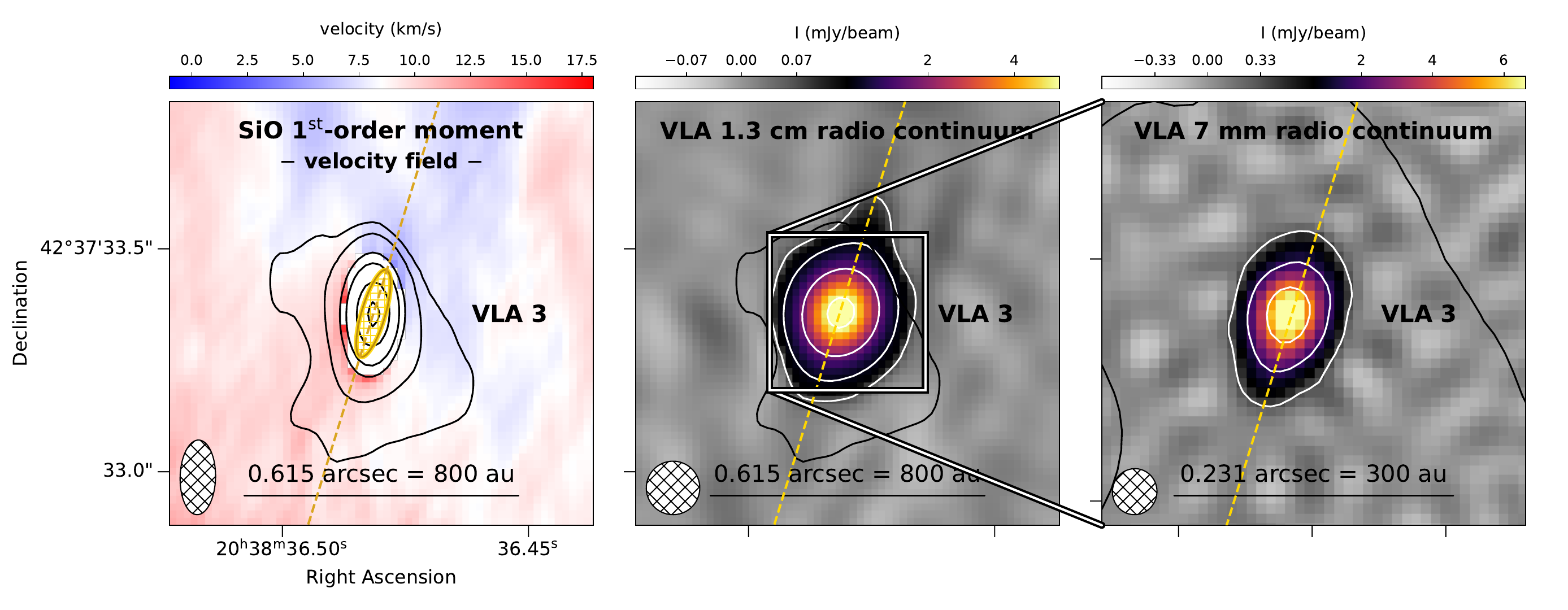}
\vspace{-0.5cm}
\caption{(\textit{left panel}): SiO\,(5--4) velocity field (1$^\mathrm{st}$ order moment map) towards W75N(B)-VLA3. Black contours trace the ALMA 1.3~mm continuum emission (see Fig.~\ref{fig:continuum}). The central yellow-orange ellipse depicts the deconvolved size ($0\farcs21\times0\farcs07$) and orientation (PA $=157\pm3^\circ$) of the thermal radio jet reported by \citet{CarrascoGonzalez2010a} at 3.6~cm. The orientation of the jet at 3.6 cm (PA $=157^\circ$) is highlighted with the dashed yellow-orange line that crosses the panel from southeast to northwest. This orientation is in agreement with the tentative red to blue-shifted velocity gradient observed in the SiO\,(5--4) line. (\textit{central and right panels}): VLA 1.3~cm and 7~mm continuum emission from VLA3. Data from \citet{CarrascoGonzalez2015} and \citet{RodriguezKamenetzky2020}. White contours show the 1.3~cm continuum emission at levels 0.067, 0.34, 1.7, and 5~mJy~beam$^{-1}$ (with rms $\approx6.7$~$\mu$Jy~beam$^{-1}$ and beam $=0\farcs12$) in the central panel, and the 7~mm continuum emission at levels 0.33, 1.7, and 5~mJy~beam$^{-1}$ (with rms $\approx66$~$\mu$Jy~beam$^{-1}$ and beam $=0\farcs038$) in the right panel. The images have been generated with circular beams to emphasize the elongation of the thermal radio jet emission at 1.3~cm (deconvolved size = $0\farcs09\times0\farcs04$; PA $=162\pm1^\circ$) and 7 mm (deconvolved size = $0\farcs05\times0\farcs03$; PA $=165\pm2^\circ$), which is consistent with the orientation derived by \citet{CarrascoGonzalez2010a} and highlighted with the light-yellow dashed line. See also the study of the dependence of the thermal radio jet's size on frequency carried out by \citet{RodriguezKamenetzky2020}}
\label{fig:jet}
\end{figure*}

\begin{figure*}
\centering
\includegraphics[width=1.0\textwidth]{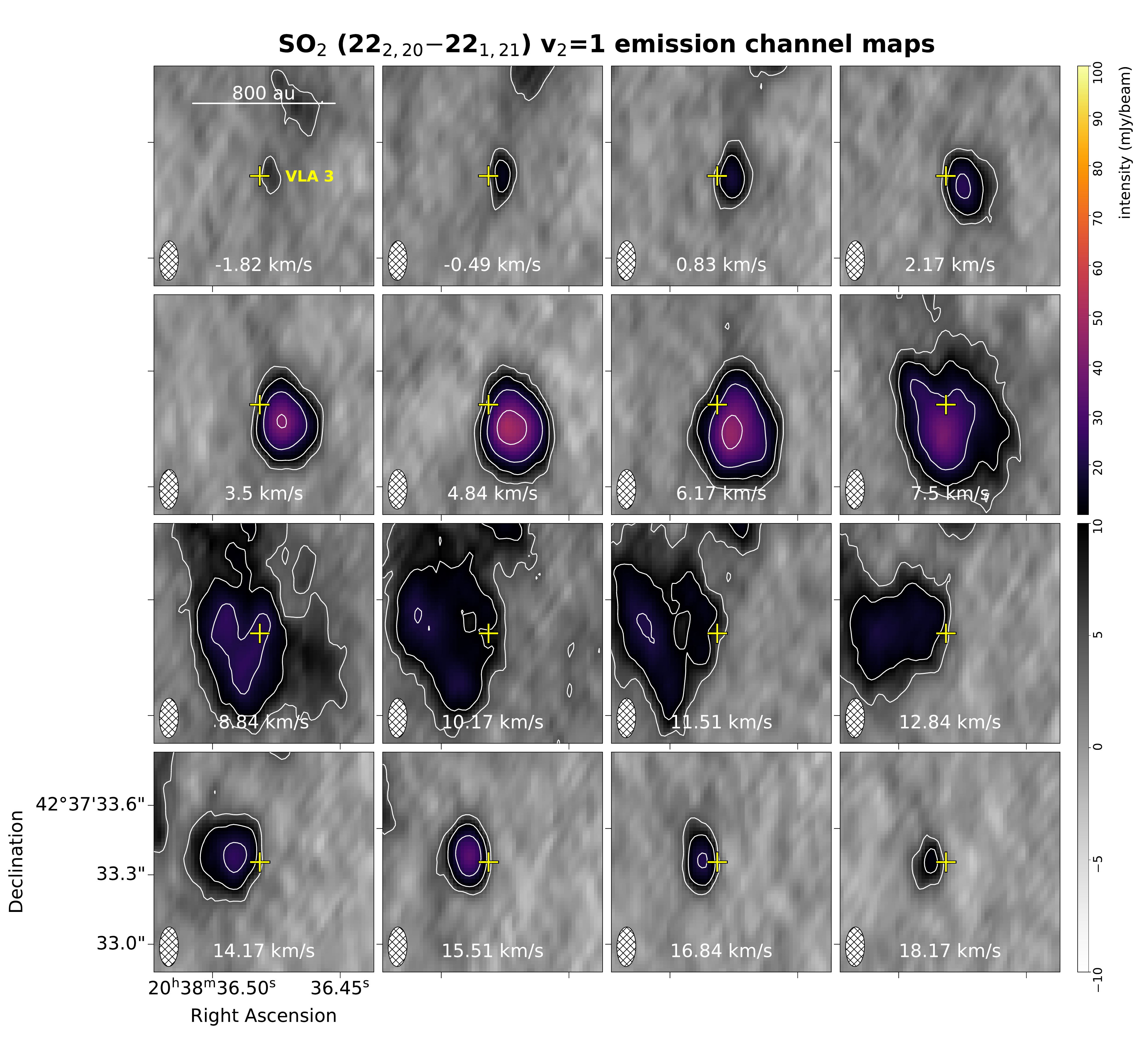}
\vspace{-0.5cm}
\caption{ALMA SO$_2$\,(22$_{2,20}$--22$_{1,21}$; v$_2$=1) emission channel maps. From top-left to bottom-right the velocity of each channel, indicated at the bottom of each panel, increases from $-1.82$~km~s$^{-1}$ to $+18.17$~km~s$^{-1}$, in steps of $\approx1.35$~km~s$^{-1}$. The solid contours mark the SO$_2$ emission at the levels 4, 8, 16, and 32 times the rms of the image ($\mathrm{rms}=1.25$~mJy~beam$^{-1}$ in channels of $\approx1.35$~km~s$^{-1}$ width; beam $=0\farcs171\times0\farcs082$, PA$=-1\fdg5$, depicted with a hatched ellipse in each panel). The yellow cross marks the position of the VLA3 YSO. The SO$_2$ emission shifts from southwest with respect to VLA3 at blue-shifted velocities to northeast at red-shifted velocities (systemic velocity at V$_\mathrm{LSR}$$\simeq$ 8.9~km~s$^{-1}$, see Sect.~\ref{sec:discussion}).}
\label{fig:SO2_channels}
\end{figure*}

%
\subsubsection{Shock and outflow tracers: SiO\,(5--4)}\label{sec:shocked-gas}

The SiO molecule has long been considered a potential shock tracer of outflows \citep[e.g.,][]{Schilke1997, Gusdorf2008, SanchezMonge2013b}. In this study, we aim to investigate the outflow associated with the thermal radio jet, given the previous detection of SiO\,(1--0) with the VLA \citep[][]{CarrascoGonzalez2010a}. Our ALMA observations of the SiO\,(5--4) line do not reveal any clear structure reminiscent of a typical bipolar outflow associated with VLA3. Instead, as previously reported by \cite{Gomez2023}, a dominant absorption feature is detected in SiO\,(5--4) within the velocity range V$_{\rm LSR}$ $\simeq$ 4--8~km~s$^{-1}$, coinciding with the brightest and most compact continuum emission of VLA3 (see the SiO velocity channel and moment maps in Figs.~\ref{fig:SiO_channels} and \ref{fig:SiO_moments}). Interestingly, the peak velocity of this absorption feature (V$_{\rm LSR}$ $\simeq$ 6.3~km~s$^{-1}$) is blue-shifted by $\sim$2.5~km~s$^{-1}$ with respect to the velocity of the ambient dense gas, as estimated by the kinematic analysis presented in Sect.~\ref{sec:discussion}. A strong absorption feature is also observed in H$_2$CO within the same velocities as in SiO and coincident with the compact continuum emission (see Sect.~\ref{sec:dense-gas}).

The SiO velocity field derived from the first-order moment map (see left panel of Fig.~\ref{fig:jet}) reveals a tentative velocity gradient with blue-shifted velocities to the northwest and red-shifted velocities towards the southeast. We note that with the current data, we can not rule out the possibility that the blue-shifted emission might be partially associated with extended SiO emission from VLA2 (see Fig.~\ref{fig:SiO_moments}). Additionally, the presence of the strong absorption feature prevents us from conducting a more detailed analysis of the SiO kinematics in the near vicinity of VLA3. Interestingly, the orientation of this tentative SiO velocity gradient is consistent with the elongation of the thermal radio jet (PA $=157\pm3^\circ$) reported by \citet{CarrascoGonzalez2010a} at 3.6~cm (see yellow ellipse and dashed line in Fig.~\ref{fig:jet}). The central and right panels of Fig.~\ref{fig:jet} show the radio continuum emission from the radio jet at 1.3~cm and 7~mm observed with the VLA at higher angular resolution \citep[data from][]{CarrascoGonzalez2015, RodriguezKamenetzky2020}. These two images have been generated with circular beams ($0\farcs12$ at 1.3~cm, and $0\farcs038$ at 7~mm) resulting in an angular resolution comparable or better than the ALMA observations presented here. The radio continuum emission at both 1.3~cm and 7~mm reveal an elongation in the southeast to northwest direction in agreement with the orientation derived by \citet{CarrascoGonzalez2010a} and with the tentative SiO velocity gradient. In summary, the SiO\,(5--4) observations together with the radio continuum emission maps support the presence of a jet powered by VLA3, at an orientation PA $\approx157^\circ$.

\begin{figure*}
\centering
\includegraphics[width=1.0\textwidth]{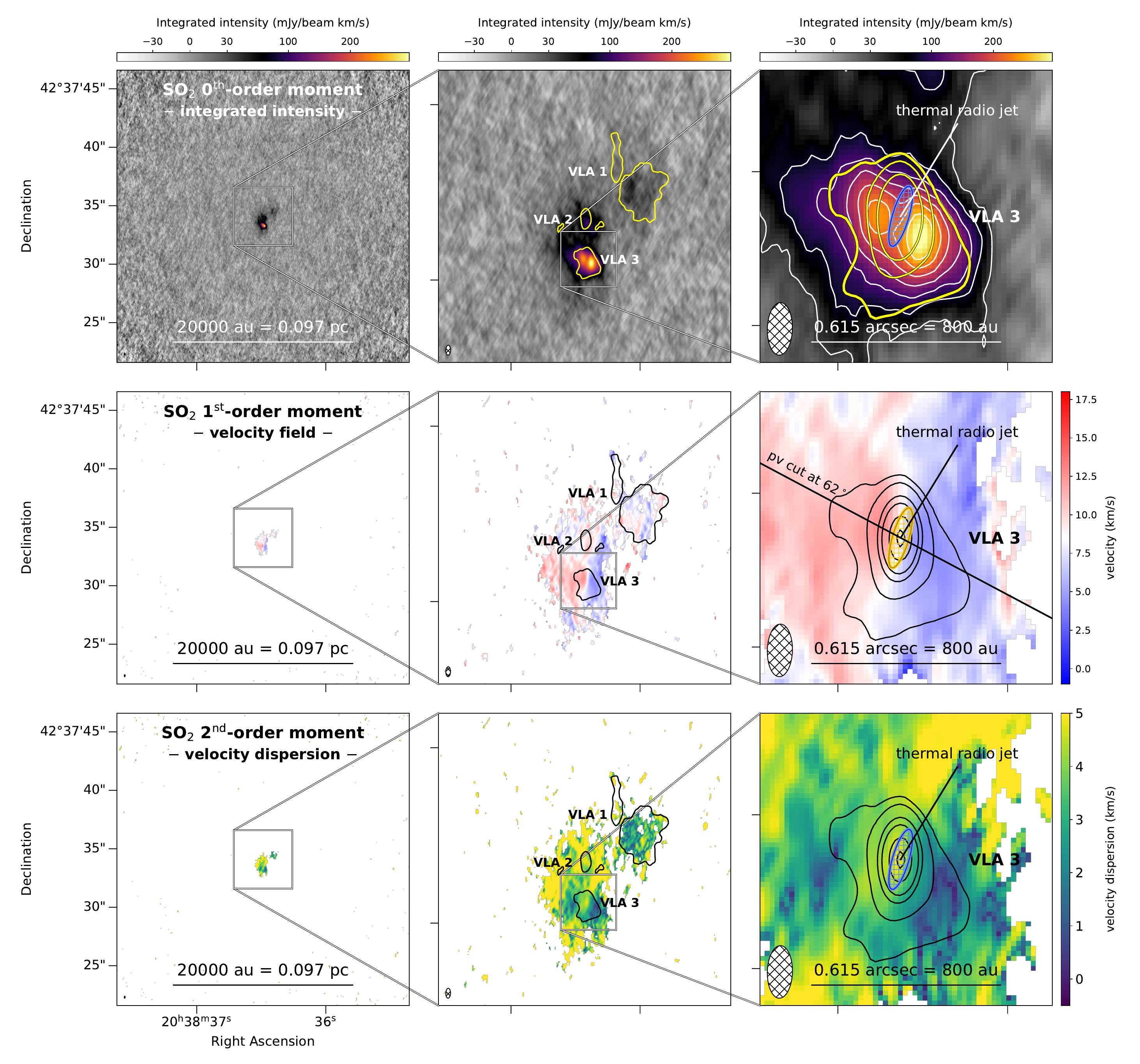}
\vspace{-0.5cm}
\caption{SO$_2$\,(22$_{2,20}$--22$_{1,21}$; v$_2$=1) moment maps. Rows from top to bottom show: integrated intensity (0$^\mathrm{th}$ order); velocity field (1$^\mathrm{st}$ order); and velocity dispersion (2$^\mathrm{nd}$ order). The velocity interval used to generate the moment maps ranges from $-1.82$ up to $+18.17$~km~s$^{-1}$ (see Fig.~\ref{fig:SO2_channels}). Columns from left to right show different zoomed-in views of the W75N(B) region, as described in Fig.~\ref{fig:continuum}. The ALMA 1.3~mm continuum emission (see Fig.~\ref{fig:continuum}) is shown in yellow contours in the top-row panels, and in black contours in the middle- and bottom-row panels. White contours in top-right panel shows the SO$_2$ integrated emission at levels 5, 10, 15, 20, 25, 30, and 35 times 7.4~mJy~beam$^{-1}$~km~s$^{-1}$, with the beam depicted by hatched ellipses. The blue and yellow ellipse in the right-column panels depicts the size and orientation of the thermal radio jet reported by \citet{CarrascoGonzalez2010a}. The diagonal solid black line in the right-column, middle-row panel shows the direction of the position-velocity (PV) cut, with PA$=62^\circ$, shown in Fig.~\ref{fig:PVcut}.}
\label{fig:SO2_moments}
\end{figure*}

\begin{figure*}
\centering
\includegraphics[width=0.95\textwidth]{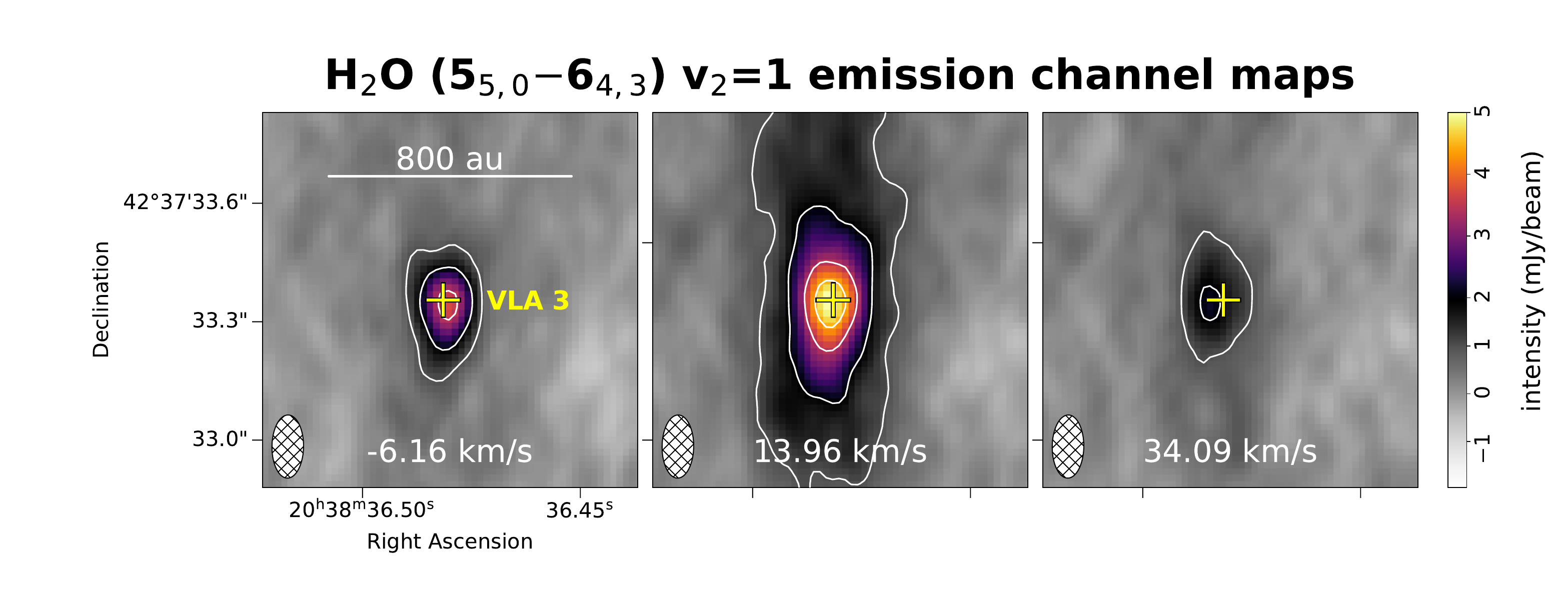}
\vspace{-0.5cm}
\caption{ALMA H$_2$O\,(5$_{5,0}$--6$_{4,3}$; v$_2$=1) emission channel maps. From left to right the velocity of each channel, indicated at the bottom of each panel, increases from $-6.16$~km~s$^{-1}$ to $+34.09$~km~s$^{-1}$, in steps of $\approx20$~km~s$^{-1}$. The solid contours mark the H$_2$O emission at the levels 4, 8, 14, and 18 times the rms of the image ($\mathrm{rms}=0.25$~mJy~beam$^{-1}$ in channels of $\approx20$~km~s$^{-1}$ width; beam $=0\farcs171\times0\farcs082$, PA$=-1\fdg5$, depicted with a hatched ellipse in each panel). The yellow cross marks the position of the VLA3 YSO. Despite the coarse spectral resolution, the H$_2$O emission shifts from west with respect to VLA3 at blue-shifted velocities to east at red-shifted velocities, similar to what is observed in other dense gas tracers (see e.g., Fig.~\ref{fig:SO2_channels}).}
\label{fig:H2O_channels}
\end{figure*}

%
\subsubsection{Dense gas tracers: SO$_2$, H$_2$CO, HC$_3$N, HNCO, CH$_3$CN, H$_2$O}\label{sec:dense-gas}

Figure~\ref{fig:SO2_channels} displays velocity channel maps showing the spatial distribution of the SO$_2$~($22_{2,20}$--$22_{1,21}$ ; v$_2$=1) emission. These maps cover the velocity range V$_{\rm LSR}$ = $-1.82$~km~s$^{-1}$ to $+18.17$~km~s$^{-1}$, with a channel spacing of $\approx$1.3~km~s$^{-1}$. The location of VLA3 is marked with a yellow cross in each velocity panel for reference. A distinct kinematic trend is evident in the SO$_2$ emission: At velocities of $\simeq$ $-1.82$ to $+6$~km~s$^{-1}$, the emission peak in each channel is located southwest of VLA3. As the velocity increases towards higher values, from $\simeq$ $+8$~km~s$^{-1}$ to $+18$~km~s$^{-1}$, the emission peak systematically shifts to the northeast of VLA3. This velocity gradient, perpendicular to the orientation of the thermal radio jet and SiO velocity gradient, suggests organised gas motions probably related to the rotation of dense gas around the massive protostar (see Sect.~\ref{sec:discussion}).

The SO$_2$ emission is further investigated through moment maps, presented in Fig.~\ref{fig:SO2_moments}. The figure displays the integrated intensity ($0^{\rm th}$ order moment: MOM0), velocity field ($1^{\rm st}$ order moment: MOM1), and velocity dispersion ($2^{\rm nd}$ order moment: MOM2) of the SO$_2$ emission in separate rows. The integrated intensity map of SO$_2$ (MOM0, Fig.~\ref{fig:SO2_moments} top-row) reveals a compact, elongated structure oriented in a northeast to southwest direction (PA $\simeq 62^{\circ}$), spanning $\sim$$0\farcs7$ ($\sim$$900$~au at a 3$\sigma$ level, slightly larger than the extent of the mm continuum emission, see Sect.~\ref{sec:continuum}) and exhibiting two intensity peaks separated by $\sim$$0\farcs25$ ($\sim$$300$~au), with the southern one being slightly brighter. A 2D Gaussian fit to the SO$_2$ integrated intensity map results in a deconvolved FWHM of $0\farcs51\pm0\farcs04\times0\farcs34\pm0\farcs03$, with PA $=58\pm9^\circ$. The blue ellipse overlaid on the panel indicates the deconvolved size ($0\farcs21\times0\farcs07$) and orientation (PA $=157^\circ$) of the VLA3 thermal radio jet previously reported by \citet{CarrascoGonzalez2010a} at 3.6~cm (see also Fig.~\ref{fig:jet}). The thermal radio jet is situated between these two SO$_2$ emission maxima, and displays an orientation perpendicular to the elongated structure of the gas emission. The velocity field of the SO$_2$ emission, as depicted by the first-order moment (Fig~\ref{fig:SO2_moments} middle-row), shows a velocity gradient extending along the elongated structure. This velocity variation was identified earlier through the velocity channel maps shown in Fig.~\ref{fig:SO2_channels}. Conversely, the second-order moment map (Fig.~\ref{fig:SO2_moments} bottom-row) reveals a velocity dispersion of $\sim$3--4~km~s$^{-1}$, with no significant variation observed along the molecular structure.

The emission from other dense gas tracer molecules identified towards VLA3 (H$_2$CO, HC$_3$N, HNCO, CH$_3$CN; transitions listed in Table~\ref{tab:spectral-lines}) displays a spatial and kinematic distribution which in general is similar to that observed in SO$_2$ (see Appendix~\ref{app:additional}, and Figs.~\ref{fig:H2CO_channels} to \ref{fig:CH3CN_moments}), and therefore is likely associated with intrinsic properties of the circumstellar dense gas around the VLA3 protostar. However, these lines, which have lower excitation energies (see Table~\ref{tab:spectral-lines}), exhibit a more extended distribution across the systemic velocity (V$_\mathrm{LSR}\simeq$~8.9~km~s$^{-1}$) range of VLA3, likely tracing the surrounding local environment. Yet, their extreme velocity channels distinctly reveal, as in SO$_2$, a velocity gradient oriented southwest to northeast, suggestive of rotation motions.

In particular, the H$_2$CO emission towards VLA3 is detected across a broad velocity range, from V$_{\rm LSR}$ $\simeq -1.7$ to $+18$~km~s$^{-1}$ (see Fig.~\ref{fig:H2CO_channels}) similar to that of the SO$_2$ emission (Fig.~\ref{fig:SO2_channels}). The integrated intensity (MOM0) of H$_2$CO reveals an elongated structure oriented northeast to southwest, featuring two intensity peaks and a velocity gradient comparable to that observed in SO$_2$ (compare Figs.~\ref{fig:H2CO_moments} and \ref{fig:SO2_moments}). Nevertheless, this velocity gradient is partially obscured by intense extended H$_2$CO emission spanning the entire W75N(B) region within V$_{\rm LSR} \simeq +5$ to $+13$~km~s$^{-1}$ (see Fig.~\ref{fig:H2CO_channels}, and \citealt{Gomez2023}), and by a strong absorption feature at V$_{\rm LSR} \simeq +6$~km~s$^{-1}$ (see Fig.~\ref{fig:H2CO_channels}), analogous to the SiO absorption feature. The blue-shifted absorption seen in the SiO and H$_2$CO channel maps is reminscent of the classical P-Cygni profiles and may suggest the presence of outflowing or expanding motions in the molecular gas around VLA3. We note that the peak brightness temperature of the 1.3~mm continuum emission is $\approx100$~K, which is higher than the upper energy levels of the SiO and H$_2$CO transitions (about 20--30~K, cf.\ Table~\ref{tab:spectral-lines}) but lower than the energy levels of the other molecular lines studied in this work (between 130 and 3500~K). Higher spatial resolution observations, capable of resolving the most compact structures, may reveal a much brighter continuum source against which other molecular species might also show self-absorbed profiles.

We also detected the vibrationally excited H$_2$O\,(5$_{5,0}$--6$_{4,3}$ ; $v_2$=1) transition at 232.6867~GHz (see Fig.~\ref{fig:spectra} and Table~\ref{tab:spectral-lines}). Vibrationally excited H$_2$O lines have been identified as excellent tracers of the inermost, and hottest regions of discs surrounding (massive) YSOs \citep[see e.g.,][]{Ginsburg2018, Maud2019}. Unfortunately, this transition is covered in our setup within the spectral window spw\,0 with a low spectral resolution ($\approx20$~km~s$^{-1}$). Figure~\ref{fig:H2O_channels} shows the H$_2$O emission map through the three velocity channels at which we detect emission from the water line. The emission is compact and spatially coincident with the bright 1.3~mm continuum source, with the emission in the central velocity channel slightly elongated in the north-south direction (deconvolved size $0\farcs89\times0\farcs29$, with PA $=0^\circ$). Interestingly, and despite the coarse spectral resolution, the H$_2$O emission peak shifts from west to east with respect to VLA3 when moving from blue- to red-shifted velocities. This is consistent with the shift observed for the other dense gas tracers (see e.g., Fig.~\ref{fig:H2CO_channels} and figures in Appendix~\ref{app:additional}), and supports the idea that this H$_2$O vibrational line may also trace rotational motions around VLA3 as found in other massive protostars. However, higher spatial and spectral resolution observations are necessary to properly characterize the structure and kinematics traced by the H$_2$O line.

\begin{figure}
\centering
\includegraphics[width=1.0\columnwidth]{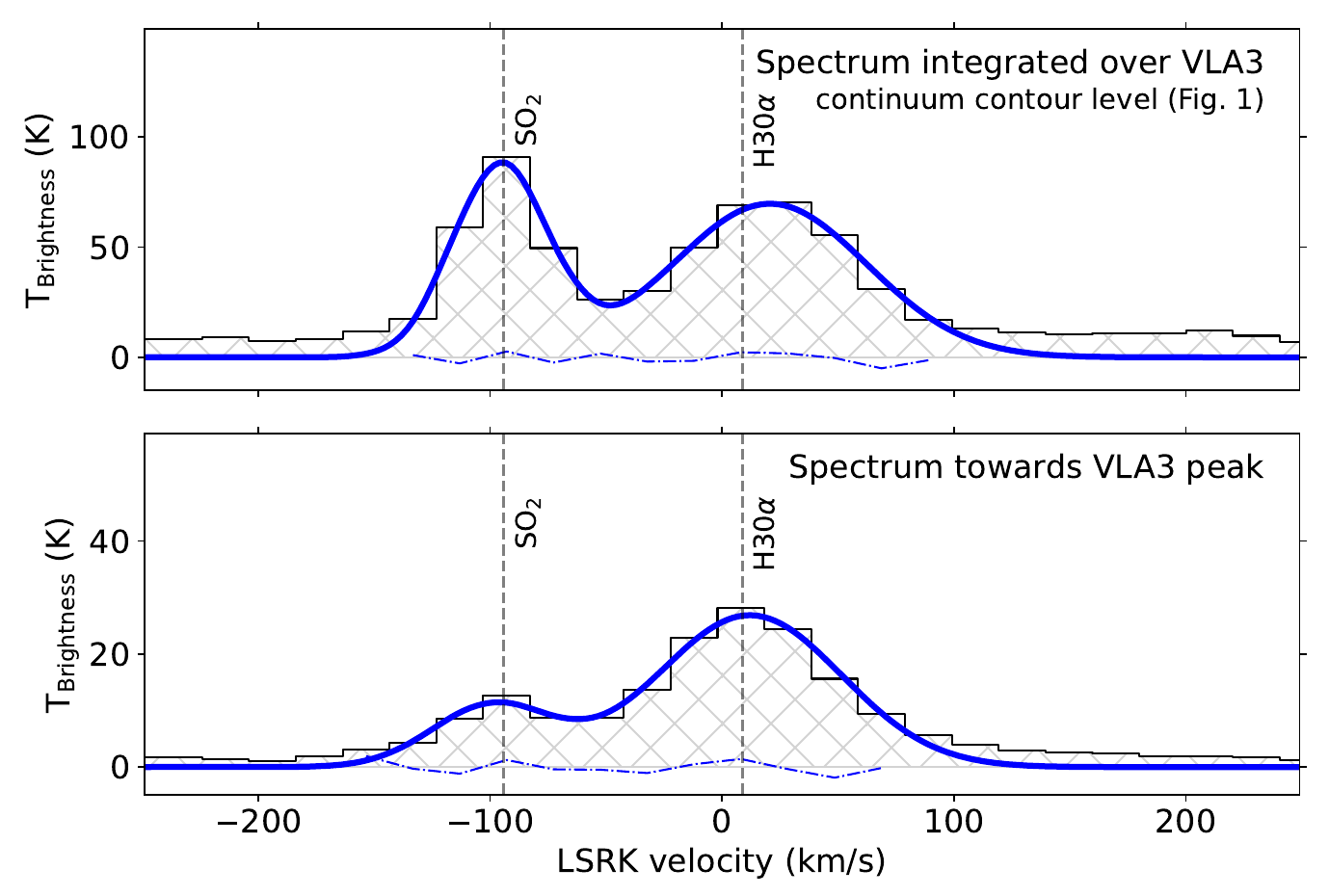}
\vspace{-0.5cm}
\caption{ALMA spectrum around the H30$\alpha$ line (\textit{top panel}) integrated over the region defined by the 3~mJy~beam$^{-1}$ contour level of the 1.3~mm continuum emission (see Fig.~\ref{fig:continuum}), and (\textit{bottom panel}) extracted towards the continuum peak of VLA3. The observational data are shown with histograms. The fitted spectra are shown with a blue, solid line (see Sect.~\ref{sec:ionized-gas}), with the residuals depicted with a blue, dot-dashed line. The vertical dashed lines mark the location of the H30$\alpha$ (at rest frequency 231.901~GHz) and SO$_2$\,(14$_{3,11}$--14$_{2,12}$ ; v$_2$=1) (at 231.980~GHz) lines, for the systemic velocity of VLA3 (i.e., 8.9~km~s$^{-1}$; see Sect.~\ref{sec:pvplots}).}
\label{fig:H30a_fit}
\end{figure}

%
\subsubsection{Ionised gas tracers: H30$\alpha$ recombination line}\label{sec:ionized-gas}

The hydrogen radio recombination line (RRL) H30$\alpha$, which indicates the presence of ionised gas, falls within the frequency range of our observations, albeit within the spectral window spw\,0 which was optimised for continuum observations, and consequently has a lower spectral resolution ($\approx20$~km~s$^{-1}$) compared to the other spw's (see Table~\ref{tab:spectral-setup}). The spectrum integrated towards VLA3, as shown in Fig.~\ref{fig:spectra}, displays a spectral feature at about 231.9~GHz that coincides with the H30$\alpha$ frequency. We have searched for other molecular species with spectral lines near the H30$\alpha$ frequency that might account for the detected line. The only possible contaminant is $^{33}$SO$_2$, with a group of transitions (3$_{9,{\rm K_{c}}}$--2$_{10,{\rm K_{c}}}$) located between 231.89 and 231.90~GHz. However, using a typical isotopic ratio of  $^{32}$S/$^{33}$S of 88 \citep[see][]{Yan2023}, the expected contribution from this contaminant to the observed emission line would be only around 15\%, based on the intensity of the SO$_2$ lines detected within the same spectral range. Based on this, we assign the bright spectral feature at 231.9~GHz to the H30$\alpha$ line.

\begin{figure*}
\centering
\includegraphics[width=1.0\textwidth]{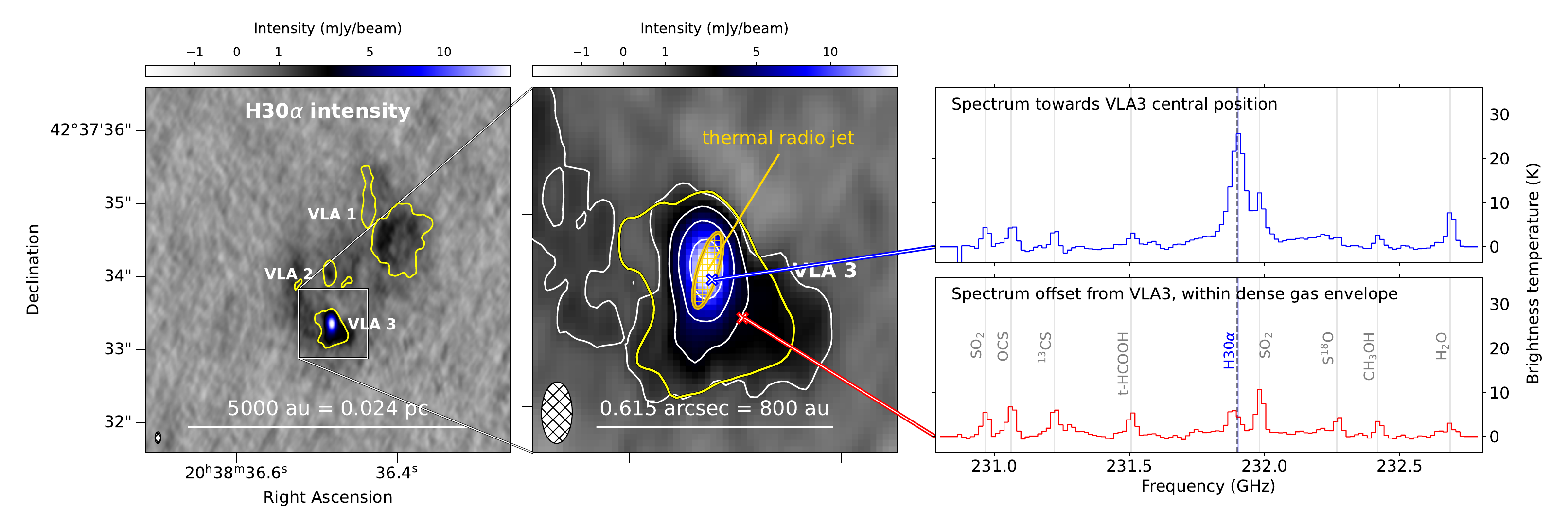}
\vspace{-0.5cm}
\caption{H30$\alpha$ images and spectra towards VLA3. Left and middle panels: intensity emission map at the channel with frequency 231.9~GHz for (\textit{left}) a larger region covering the three main YSOs VLA1, VLA2 and VLA3; and (\textit{middle}) zoomed in to VLA3. The yellow contour depicts the 3~mJy~beam$^{-1}$ level of the ALMA 1.3~mm continuum emission. White contours in the middle panel show the H30$\alpha$ emission at levels 7, 14, 28, and 56 times 0.18~mJy~beam$^{-1}$. The yellow-orange ellipse marks the location and orientation of the thermal radio jet \citep{CarrascoGonzalez2010a}. The two right panels show the spectra extracted towards two locations: the radio jet (\textit{top panel}) and a location shifted southwest by $0\farcs07$ ($\approx95$~au; \textit{bottom panel}).}
\label{fig:H30alpha}
\end{figure*}

\begin{figure}
\centering
\includegraphics[width=1.0\columnwidth]{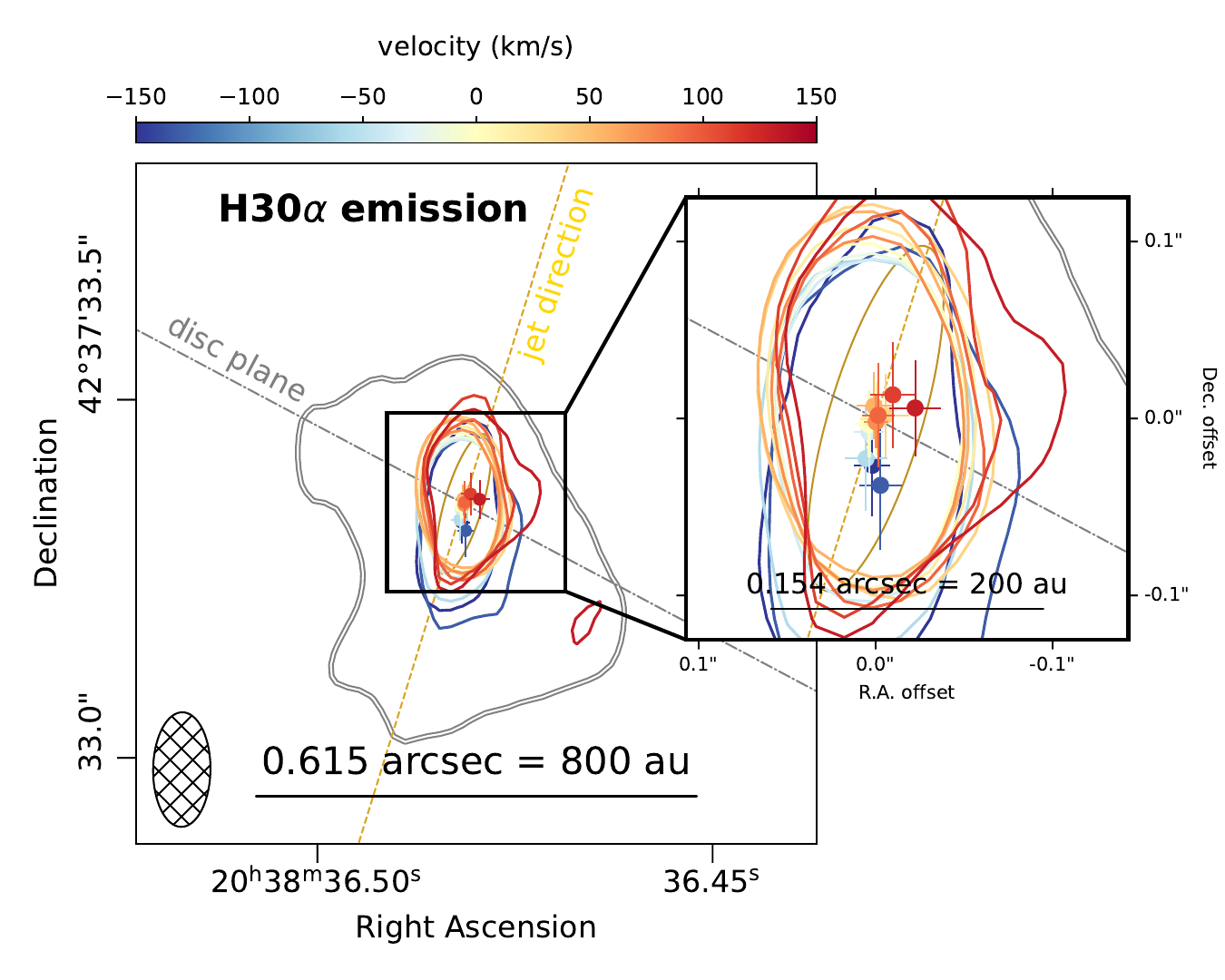}
\vspace{-0.5cm}
\caption{Distribution of the peaks of the H30$\alpha$ line emission (solid colored circles) obtained with a 2D Gaussian fit channel by channel (from Fig.~\ref{fig:H30a_channels}), together with their 1-$\sigma$ positional uncertainties (derived from the 2D Gaussian fit, and marked with colored crosses). For each peak, the corresponding 50\% contour level is also drawn (using the same color as the peak). The color corresponds to the line-of-sight velocity, according to the scale displayed on the top ranging from $-150$ to $+150$~km~s$^{-1}$. The peaks and contours corresponding to the three channels contaminated by the neighboring SO$_2$ transition (cf.\ Fig.~\ref{fig:H30a_channels}) have been excluded. The black-white contour depicts the 3~mJy~beam$^{-1}$ level of the ALMA 1.3~mm continuum emission (see Fig.~\ref{fig:continuum}). The grey, dot-dashed line indicating the disc plane corresponds to the orientation of the position-velocity cut discussed in Sect.~\ref{sec:pvplots} (see also Fig.~\ref{fig:SO2_moments}). The yellow-orange, dashed line marks the direction of the radio jet (see Fig.~\ref{fig:jet}). The synthesised beam of the H30$\alpha$ maps is shown in the bottom-left corner with a hatched ellipse. The zoom-in panel shows a closer view of the central $0\farcs25\times0\farcs25$ area.}
\label{fig:H30a_positions}
\end{figure}

Figure~\ref{fig:H30a_fit} shows a close-up view of the spectrum around the frequency of the H30$\alpha$ line. The top panel shows the spectrum integrated over the 3~mJy~beam$^{-1}$ continuum level (see Fig.~\ref{fig:continuum}) used to determined properties of the VLA3 continuum emisison, while the bottom panel shows the spectrum extracted towards the peak position of the continuum source. In order to determine the intensity and linewidth of the H30$\alpha$ line, we have fitted two Gaussians to account for the spectral line features associated with the H30$\alpha$ line itself and with the neighboring SO$_2$\,(14$_{3,11}$--4$_{2,12}$; v$_2$=1) line at 231.980~GHz. The best fit is shown in blue in Fig.~\ref{fig:H30a_fit}. The fit to the integrated spectrum delivers for the H30$\alpha$ line a peak intensity of $70\pm5$~K (or $42\pm6$~mJy~beam$^{-1}$), centered at $20\pm4$~km~s$^{-1}$ (with respect to the H30$\alpha$ rest frequency), and with a velocity dispersion\footnote{This velocity dispersion, $\sigma$, corresponds to a linewidth or full-width at half-maximum (FWHM) of the line, $\Delta v = \sigma \cdot (8 \ln 2)^{1/2}$, equal to $98\pm10$~km~s$^{-1}$.} of $42\pm8$~km~s$^{-1}$. The Gaussian fit of the neighboring SO$_2$ feature results in a peak intensity of $87\pm7$~K, centered at $8\pm2$~km~s$^{-1}$ (with respect to the SO$_2$ rest frequency of 231.980~GHz), and with a velocity dispersion of $21\pm4$~km~s$^{-1}$ (likely severely affected by the coarse spectral resolution of $\approx20$~km~s$^{-1}$). Similarly, the fit to the spectrum extracted towards the peak results in a H30$\alpha$ line intensity of $27\pm2$~K, centered at $12\pm2$~km/s, and with a velocity dispersion of $40\pm6$~km~s$^{-1}$. The fit parameters for the SO$_2$ line are an intensity of $11\pm2$~K, centered at $4\pm4$~km~s$^{-1}$, with a dispersion of $26\pm11$~km~s$^{-1}$. The spectra shown in Fig.~\ref{fig:H30a_fit} shows a tentative departure from Gaussianity at high velocities, likely tracing high-velocity wings. This high-velocity component is also seen in the channel emission maps shown in Fig.~\ref{fig:H30a_channels}. In this figure, the emission is dominated by a compact, bright component associated with the H30$\alpha$ line and with its peak at the velocity channel 8.16~km~s$^{-1}$ (consistent with the systemic velocity of VLA3). This component is surrounded by an extended structure, which is brightest at the velocity channel 28.35~km~s$^{-1}$ and is likely related to the blended spectral transitions of $^{33}$SO$_2$. The compact component that traces the H30$\alpha$ line emission extends up to high velocities of $\pm150$~km~s$^{-1}$.

\begin{figure*}
\centering
\includegraphics[width=1.0\textwidth]{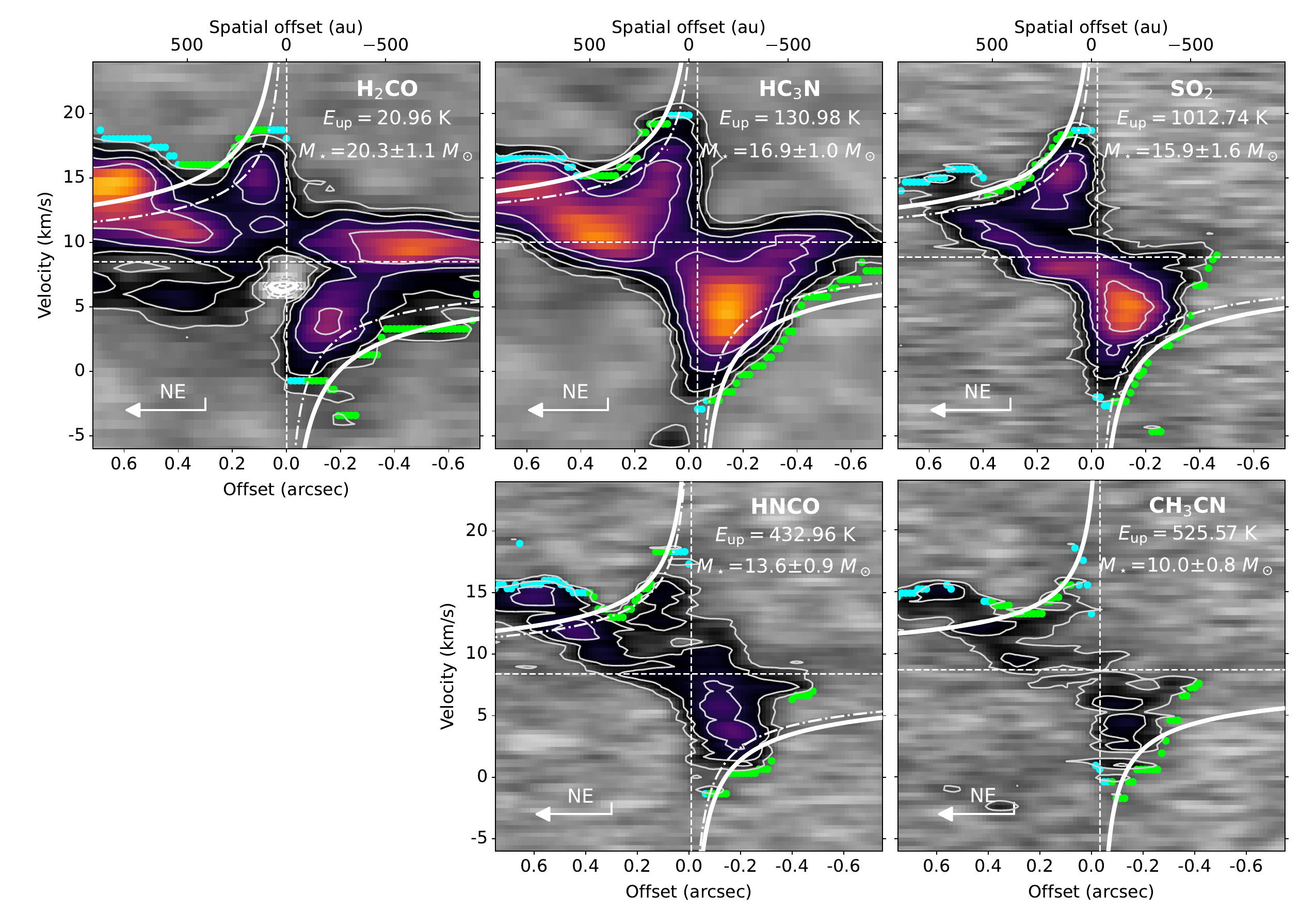}
\vspace{-0.5cm}
\caption{Position-velocity (pv) plots for H$_2$CO, HC$_3$N, SO$_2$, HNCO, and CH$_3$CN, centered at VLA3 and in the direction PA=62$^\circ$ (see direction of the pv-cut in the velocity field panel of Fig.~\ref{fig:SO2_moments}). The color maps show the emission along the pv plots which has been averaged over nine pixels (of $0\farcs016$ each) across the direction of the pv cut. White dashed and solid contours show levels $-16$, $-8$, $-4$, 4, 8, 16, 32 times the rms of the images. We derive the rms from the median absolute deviation of the pv plots, resulting in values of 0.88, 0.93, 0.95, 1.04 and 0.99~mJy~beam$^{-1}$ for H$_2$CO, HC$_3$N, SO$_2$, HNCO, and CH$_3$CN, respectively. The thicker solid white lines mark the best fit Keplerian rotation models for each of the molecular species, with the fitted masses listed in the top-right corner of each panel (see Sect.~\ref{sec:pvplots} and Table~\ref{tab:KeplerianFit} for more details). For comparison, the Keplerian rotation curve for a central mass of 10~$M_\odot$ is shown with dot-dashed white curves. Lime and cyan symbols mark the highest velocities at 4$\times$rms intensity levels for different radial offsets, with the lime symbols being those used to fit the Keplerian disc-like rotating model \citep[see][and Sect.~\ref{sec:pvplots}]{Seifried2016}.}
\label{fig:PVcut}
\end{figure*}

In the left panels of Fig.~\ref{fig:H30alpha}, we present the emission of the H30$\alpha$ line, overlaid with the 1.3~mm continuum emission contours and the radio jet location identified by \citet{CarrascoGonzalez2010a}. The peak of the H30$\alpha$ emission is spatially coincident with the 1.3~mm continuum peak and the radio jet, although there is also weaker emission distributed across the map, particularly towards VLA1. The weaker, extended emission may be related to other molecular species that are blended with the H30$\alpha$ line due to the limited spectral resolution. The right panels of Fig.~\ref{fig:H30alpha} depict the spectra extracted at the position of the radio jet (top) and at a location shifted southwest by $0\farcs07$ ($\approx95$~au; bottom). The broad spectral line feature corresponding to H30$\alpha$ is most pronounced at the radio jet's position, nearly vanishing when slightly offset from it, thereby supporting its spatial association with the VLA3 jet and mm continuum peak. Additionally, the consistent intensities of other spectral lines across both positions (cf.\ Fig.~\ref{fig:H30alpha}) further support the H30$\alpha$ association with the jet or the central region of VLA3, since these other lines are associated with molecular species that trace the whole extent of the dense core and not only the inner region. It is noteworthy that the lines from other molecular species are considerably weaker than the H30$\alpha$ line at the location of the radio jet, with only the vibrationally excited H$_2$O transition at 232.868~GHz remaining prominent (see Sect.~\ref{sec:dense-gas} and Fig.~\ref{fig:H2O_channels}). However, and contrary to the H$_2$O line, the H30$\alpha$ channel maps shown in Fig.~\ref{fig:H30a_channels} do not reveal a clear spatial shift of the peak emission in the east-west direction. Interestingly, a tentative north-south spatial shift, coinciding with the radio jet direction, is observed instead. This tentative spatial shift is shown in Fig.~\ref{fig:H30a_positions}, where the colored circles depict the H30$\alpha$ peak position for each velocity channel, with the colors corresponding to the channel velocity (ranging from $-150$ to $+150$~km~s$^{-1}$). The north-south spatial shift also shows a tentative velocity gradient (with red- and blue-shifted velocities located to the north and south of VLA3, respectively), which, if real, would be opposite to the velocity gradient seen in SiO (see Fig.~\ref{fig:jet}). We note that spatial overlap of blue and red-shifted gas can occur in outflows observed close the plane of the sky \citep[see e.g.,][for the case of DR21]{CruzGonzalez2007, Skretas2023}, as might be the case for the jet/outflow in VLA3. However, we note that the tentative spatial shift and velocity gradient observed in the H30$\alpha$ line must be taken cautiously, as the limited angular resolution of the observations, especially in the north-south direction, hinders a more precise confirmation. In Sect.~\ref{sec:RRLs}, we discuss the properties of the H30$\alpha$ emission and its potential connection to the radio jet powered by VLA3.

%
\section{Discussion}\label{sec:discussion}

%
\subsection{A compact, rotating disc around W75N(B)-VLA3}\label{sec:pvplots}

To comprehensively analyse the spatial and kinematic distribution of the dense gas tracers discussed in Sect.~\ref{sec:dense-gas}, we constructed the position-velocity (PV) diagrams shown in Fig.~\ref{fig:PVcut}. The position angle of the PV cut direction is PA $=62^\circ$, and is determined by the line that connects the two intensity maxima observed in the integrated intensity (MOM0) images of different dense gas tracers (see e.g., Fig.~\ref{fig:SO2_moments}). This orientation is in agreement with the direction of the velocity gradient observed in the velocity field (MOM1) maps (see e.g., Figs.~\ref{fig:SO2_moments} and \ref{fig:H2CO_moments}). Furthermore, the PV cut is almost perpendicular to the direction of the thermal radio jet (PA $=157\pm3^\circ$; \citealt{CarrascoGonzalez2010a}), thus favoring the interpretation of the observed velocity gradient tracing rotation motions perpendicular to the jet direction. The center of the PV cut corresponds to the location of VLA3, with coordinates $\alpha(\mathrm{J2000})=20^\mathrm{h}38^\mathrm{m}36.4815^\mathrm{s}$ and $\delta(\mathrm{J2000})=+42^{\circ}37^{\prime}33.355^{\prime\prime}$. The PV diagrams shown in Fig.~\ref{fig:PVcut} were generated averaging over nine pixels (of $0\farcs016$ each) across the direction of the cut to enhance the sensitivity.

\begin{table}
\centering
\caption{\label{tab:KeplerianFit}Fitted parameters of the Keplerian-rotation models shown in Fig.~\ref{fig:PVcut}}
\begin{tabular}{l c c c c}
\hline\hline 
  
& $M_{*,\mathrm{fit}}$
& $v_0$
& $r_0$
& $r_0$
\\
  Species
& ($M_\odot$)
& (km~s$^{-1}$)
& (au)
& (arcsec)
\\
\hline 
H$_2$CO  & $20.3\pm1.2$ & \phn$8.5\pm0.3$ & \phn$-1\pm14$ & \phn$-0.001\pm0.010$ \\
HC$_3$N  & $16.9\pm1.0$ & $10.0\pm0.3$    & $-41\pm11$    & \phn$-0.032\pm0.008$ \\
SO$_2$   & $15.9\pm1.6$ & \phn$8.9\pm0.5$ & $-29\pm22$    & \phn$-0.022\pm0.017$ \\
HNCO     & $13.6\pm1.0$ & \phn$8.4\pm0.4$ & $-12\pm19$    & \phn$-0.009\pm0.015$ \\
CH$_3$CN & $10.0\pm0.9$ & \phn$8.7\pm0.3$ & $-44\pm15$    & \phn$-0.034\pm0.012$ \\
\hline
\end{tabular}
\begin{tablenotes}
\footnotesize
\item {\bf Notes}. The fitted parameters are the central stellar mass $M_{*,\mathrm{fit}}=M_* \sin^2{i}$ (in $M_\odot$), the systemic velocity or velocity shift $v_0$ (in km~s$^{-1}$), and the positional shift $r_0$ (in au and arcsec) with respect to the center of the PV-cut or location of VLA3, $\alpha(\mathrm{J2000})=20^\mathrm{h}38^\mathrm{m}36.^{\!\mathrm{s}}4815$ and $\delta(\mathrm{J2000})=+42^{\circ}37^{\prime}33.^{\!\!\prime\prime}355$.
\end{tablenotes}
\end{table}

We have used the approach described in \citet[][see also the implementation by \citealt{Bosco2019}, and the \textsc{KeplerFit} Python-based package\footnote{\textsc{KeplerFit}, https://github.com/felixbosco/KeplerFit}]{Seifried2016} to fit a Keplerian disc-like rotating model to the observed PV diagrams. The white, solid lines in Fig.~\ref{fig:PVcut} show the best fits of the Keplerian rotation around a central object of mass $M_{*}$. This mass can be determined from the velocity distribution of the material around the central object, since the highest velocity at a given radial distance $r$ to the center of mass is limited by the Kepler orbital velocity $v_\mathrm{Kepler}(r)\approx\sqrt{GM_{*}/r}$, where $G$ is the gravitational constant. Based on that, we have estimated the highest velocity at a given radial distance (i.e., the outer envelope velocity) as the velocity channel at which the emission in the PV cut lies just above a threshold of 4 times the rms noise level (i.e., 4$\sigma$). This method was found to be the most robust among those tested by \citet{Seifried2016} on synthetic data where the true mass of the central object was known. The lime and cyan symbols in the PV diagrams of Fig.~\ref{fig:PVcut} mark the determined highest velocities ($v$) at the different offsets ($r$). The lime symbols are those ($v,r$) pairs used to fit the Keplerian disc-like rotation model, while the cyan symbols are those pairs that were discarded during the fitting process. These discarded pairs correspond to the northeastern region (beyond 500~au or $0\farcs4$) which is highly contaminated by extended and bright emission likely associated with larger-scale structures around VLA3 (see Sect.~\ref{sec:morphology}), and to the inner region with radius 100~au around the central object, which is not properly resolved with the current angular resolution.

The fitted Keplerian disc-like rotation model is given by
\begin{equation}
v = \mathrm{sign}(r-r_0) \cdot \sqrt{\frac{GM_{*,\mathrm{fit}}}{|r-r_0|}}+v_0
\end{equation}
where $M_{*,\mathrm{fit}}$ is the mass of the central object, which is assumed to be larger than the disc mass \citep[see][]{Bosco2019}. The sign-function is introduced to account for the opposite velocity difference from the blue- and red-shifted emission. The parameter $v_0$ provides the velocity shift or systemic velocity of the object, while the parameter $r_0$ provides the positional shift with respect to the central position of the PV-cut (i.e., VLA3). We note that masses obtained from this method are uncertain due to the unconstrained disc axis inclination $i$ to the line of sight. The actual central protostellar mass ($M_*$) can be obtained from $M_{*,\mathrm{fit}}=M_* \sin^2{i}$ \citep[see Appendix A in][]{Bosco2019} if the inclination angle is known. In the following analysis we consider $i=90^\circ$ (i.e., edge-on\footnote{An inclination angle of 70$^\circ$ or 50$^\circ$, instead of 90$^\circ$ (edge-on), would result in a central protostellar mass about 13\% or 70\% larger, respectively.}). This is based on the obscured (radio bright) HH objects detected up to distances of 5200~au from VLA3 and with high-velocity proper motions exceeding 100~km~s$^{-1}$ \citep[][]{RodriguezKamenetzky2020}, which suggest that the outflow ejecta are most probably close to the plane of the sky, and therefore the circumstellar disc is almost edge-on. The deconvolved size of the SO$_2$ integrated intensity emission (see Sect.~\ref{sec:dense-gas}) can also be used to infer the inclination angle of the disc under the assumption that the observed ellipticity is only due to the projection of a circular disc observed at a given inclination. Based on this, we obtain $b\approx a\cos(i)$, where $a$ and $b$ are the major and minor axis of the ellipse, which results in $i=50^\circ\pm15^\circ$ for the SO$_2$ structure. We note that any departure from a infinitely thin disc would result in a lower limit to the derived inclination angle, i.e., the disc would in reality be closer to edge-on. By considering an edge-on disc ($i=90^\circ$), we set a lower limit to the actual central protostellar mass, and thus we undertake a conservative approach in the following analysis (e.g., Sect.~\ref{sec:diskproperties} and \ref{sec:RRLs}).

\begin{figure}
\centering
\includegraphics[width=1.0\columnwidth]{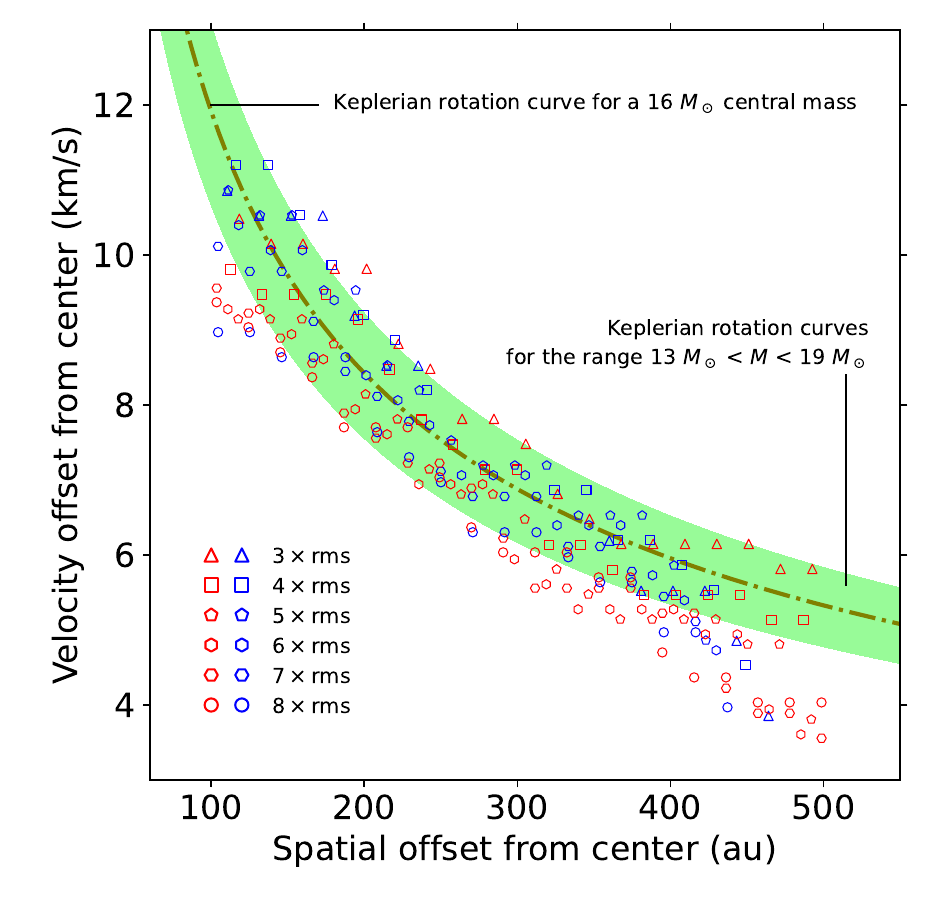}
\vspace{-0.5cm}
\caption{Radial profiles of the outer envelope velocity profile of the SO$_2$ line, extracted from Fig.~\ref{fig:PVcut}. Blue and red symbols show the blue- and red-shifted (velocity, offset) pairs, respectively, extracted at different rms thresholds (from 3 to 8 times the rms, see legend in the figure). The dot-dashed dark-green line depicts the Keplerian rotation curve for a central mass of 16~$M_\odot$. The light green shaded area depicts the range of Keplerian rotation curves for central masses in the range 13~$M_\odot$$<$$M$$<$19~$M_\odot$.}
\label{fig:pvplots_thresholds}
\end{figure}

In Table~\ref{tab:KeplerianFit} we list the fitted parameters that reproduce the Keplerian curves shown in Fig.~\ref{fig:PVcut}. The average central mass $M_{*,\mathrm{fit}}$ is $16\pm4$~$M_\odot$, after considering the five fitted spectral lines\footnote{An inclination angle of $50^\circ$ results in a central mass $M_{*}\approx27$~$M_\odot$.}. We note that the PV diagram of the vibrationally-excited SO$_2$ transition stands out as the best of the five diagrams to study the kinematics of the compact rotating structure in VLA3. This is because the H$_2$CO and HC$_2$N transitions (with energy levels $\lesssim130$~K) are more severely affected by extended emission likely not directly related to the disc, and the HNCO and CH$_3$CN transitions lack sensitivity. Based solely on the SO$_2$ transition, we also derive a central mass of $\approx16$~$M_\odot$. This mass is consistent with the stellar mass of 17--19~M$_\odot$ estimated from the correlation between the bolometric luminosity and the radio continuum luminosity at centimetre wavelengths (see Sect.~\ref{sec:continuum}). The average reference velocity $v_0$ is $8.9\pm0.6$~km~s$^{-1}$, and determines the systemic velocity of VLA3. The average offset $r_0$ is $-0\farcs019\pm0\farcs015$ (or $-25\pm19$~au) with respect to the center of the PV cut (i.e., the coordinates of VLA3), and thus, it is in agreement with the center of mass being located at the position of VLA3. In Fig.~\ref{fig:pvplots_thresholds}, we explore how the derived central mass is affected by the threshold level used when determining the outer envelope velocity in the PV diagrams \citep[see also][]{Ginsburg2018}. For this analysis, we focus on the SO$_2$ PV diagram and vary the threshold level between $3\sigma$ and $8\sigma$. The results suggest an uncertainty of about $\pm3$~$M_\odot$ on the derived central mass due to the chosen threshold level.

In summary, the dense gas tracers around VLA3 show emission consistent with a Keplerian-like rotating disc centered on an object of $\approx16$~$M_\odot$ (assuming an edge-on disc). Based on the 1.3~mm continuum emission (see Sect.~\ref{sec:continuum}), this disc-like structure has an estimated dust+gas mass (hereafter $M_\mathrm{gas}$) of 0.43--1.74~$M_\odot$ (i.e., about 10--40 times less than the central object) and extends to an estimated radius between $\approx300$~au (based on the mm continuum) and $\approx450$~au (based on the molecular lines).

%
\subsubsection{Morphology and substructures in the W75N(B)-VLA3 disc\label{sec:morphology}}

\begin{figure}
\centering
\includegraphics[width=1.0\columnwidth]{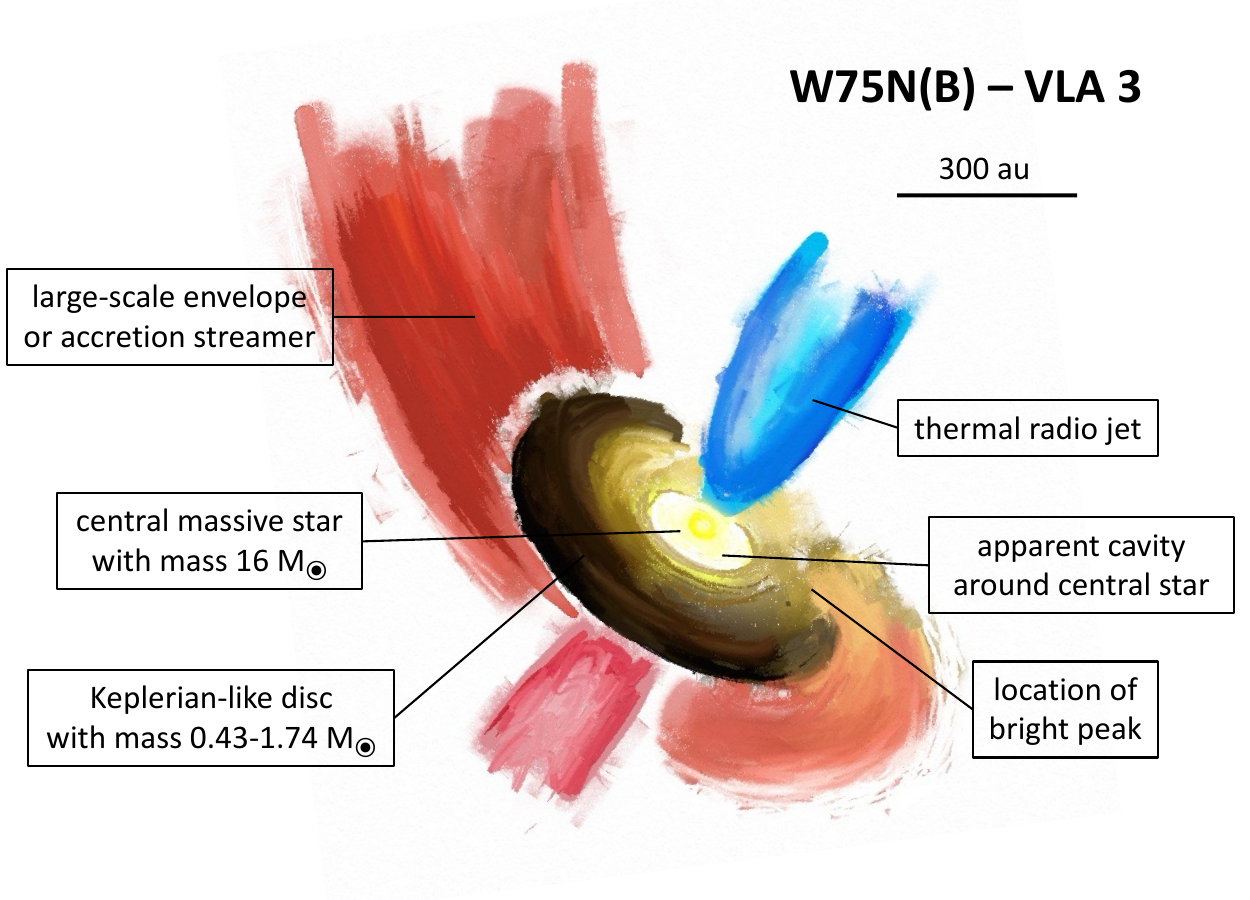}
\vspace{-0.5cm}
\caption{Artistic sketch of W75N(B)-VLA3. The central object has an estimated mass of $\approx16$~$M_\odot$ and is surrounded by a rotating disc with a gas mass of $\approx0.43$-1.74~$M_\odot$, which has a kinematic pattern consistent with Keplerian rotation (see Sect.~\ref{sec:pvplots}). The disc is oriented perpendicular to the thermal radio jet detected at centimetre wavelengths (see Sect.~\ref{sec:dense-gas}). It has an apparent molecular cavity as indicated by the detection of two peaks in various molecular species (cf.\ Figs.~\ref{fig:SO2_moments}, \ref{fig:H2CO_moments}; a possible explanation for this apparent molecular cavity is discussed in Sect.~\ref{sec:pvplots}). The brightness asymmetry of these two peaks, with the southern one being brighter, may indicate the existence of an accretion shock possibly caused by the presence of an accretion streamer as found towards other similar objects. According to the morphology of some dense gas tracers (cf.\ Fig.~\ref{fig:HC3N_moments}), this potential accretion streamer could extend towards the northwest from VLA3, where an excess of emission is seen in the integrated-intensity moment maps and PV plots (cf.\ Figs.\ref{fig:HC3N_moments} and \ref{fig:PVcut}).}
\label{fig:VLA3_sketch}
\end{figure}

As described in Sect.~\ref{sec:dense-gas}, the circumstellar disc surrounding VLA3 exhibits two peaks in the emission maps of the different dense gas tracers, with the protostar situated centrally between them and with the south-western peak being brighter  (see Fig.~\ref{fig:SO2_moments} and figures in Appendix~\ref{app:additional}). The two peaks with their different brightness suggests that the disc is not symmetrical. Such asymmetry has also been observed in other massive protostellar discs, as recently reported for Cepheus\,A~HW2 \citep[][]{Sanna2025}. In that case, the asymmetry was attributed to anisotropic accretion streams of gas and dust flowing from the outer envelope into the inner disc. In the case of W75N(B)-VLA3, these streamers are not well identified in the currently available data. However, an elongated, slightly curved structure to the north-east of VLA3 is visible in multiple species (see, e.g., moment maps of HC$_3$N and H$_2$CO in Figs.~\ref{fig:HC3N_moments} and \ref{fig:H2CO_moments}, respectively), and appears in the PV-plots shown in Fig.~\ref{fig:PVcut} as an excess of emission at an offset $>0\farcs4$ and at velocities $\sim15$~km~s$^{-1}$ (i.e., red-shifted with respect to the reference velocity of 8.9~km~s$^{-1}$). The elongated structure seen in HC$_3$N (cf.\ Fig.~\ref{fig:HC3N_moments}) is reminiscent of similar structures found towards other young stellar objects and related to accretion streamers \citep[e.g.,][]{Alves2019, Akiyama2019, Pineda2020, Garufi2022, Sanna2025}. For VLA3, and based on the red-shifted velocities of the northern elongated structure, together with the location of the brightest peak in the southern, blue-shifted region of the disc, one might imagine the existence of an anisotropic accretion streamer that brings gas from the north-eastern part of VLA3 (from the observer's perspective) and finally impacts the disc in the southern region, where the molecular emission is brighter. Figure~\ref{fig:VLA3_sketch} depicts a sketch of this hypothesis. However, a more detailed study of the extended emission around VLA3 together with new molecular observations that better trace the lower-density gas in the expected envelope or accretion streamer would be required to confirm this scenario.

Finally, as noted previously, we observed two peaks in the molecular line emission flanking the central source (cf.\ Figs.~\ref{fig:SO2_moments} and \ref{fig:H2CO_moments}). While this morphology might initially suggest a cavity devoid of molecular gas, such as one created by a toroidal gas distribution, we believe a more plausible explanation might lie in the effect of dust opacity. As shown in the upper-right panel of Fig.~\ref{fig:SO2_moments}, the continuum emission fills the region corresponding to this apparent molecular cavity. This emission, located between the two molecular line peaks, indicates the presence of dust\footnote{The presence of a real cavity, devoid of both gas and dust, would imply that the observed 1.3~mm continuum emission is not related to dust. Instead, the continuum emission would be dominated by thermal free-free emission from the radiojet. This scenario, however, appears unlikely based on the extrapolation of $\sim20$\% for the free-free contribution to the flux at 1.3~mm.} in the central region of the disc and therefore, it is expected to not be completely empty. If the continuum emission from this dust is optically thick ($\tau_{\rm dust}>1$), the molecular line emission from the same region could not be observed, since the continuum already reaches the maximum possible brightness temperature. The line emission would only be observed in the outer regions, where the dust emission is optically thin. The result is an apparent hole in molecular line emission after continuum subtraction, even though the molecular gas is present and can actually be dense. This `false' molecular cavity is naturally obtained in models that consider both line and continuum emission in circumstellar discs \citep[see, e.g., models by][]{GomezDAlessio2000}, as an effect of high dust opacity. Based on the intensity peak of the 1.3~mm continuum emission towards VLA3 (see Sect.~\ref{sec:continuum}), we estimate a beam-averaged dust opacity of $\sim$1.2 \citep[see][their Appendix~A]{Frau2010} consistent with the dust being optically thick. To confirm this interpretation for the apparent cavity, observations of the dust emission around VLA3 at different wavelengths are needed to better assess its opacity. In particular, observations at longer wavelengths are expected to mitigate the effects of dust opacity, thus enabling the study of the dust and gas distribution in the central region of VLA3. In this regard, it is noteworthy that the peak emission of the vibrationally excited H$_2$O line (see Fig.~\ref{fig:H2O_channels}) is coincident with the 1.3~mm continuum source. This suggests that the H$_2$O emission arise from very hot gas in the innermost circumstellar regions. Our limited angular precludes us from ascertaining the origin of the emission. However, as mentioned in Sect.~\ref{sec:dense-gas}, the spatial shift seen in different velocity channels suggests that part of the emission may arise from the innermost region of the circumstellar disc. The warm inner region may be mostly devoid from dust due to sublimation, so that the dust opacity does not block the line emission. Since no inner hole is seen in the continuum map, this putative dust-free area must be unresolved in our images. Moreover, part of the H$_2$O emission may also arise from the outflow, as suggested by the apparent north-south elongation of the line emission close to the direction of the jet (see Fig.~\ref{fig:H2O_channels}). Higher spatial and spectral resolution observations are necessary to shed light on the structure and morphology traced by this H$_2$O line around VLA3.

%
\subsection{Comparison with other discs/toroids around massive protostars}\label{sec:diskproperties}

\begin{figure}
\centering
\includegraphics[width=0.88\columnwidth]{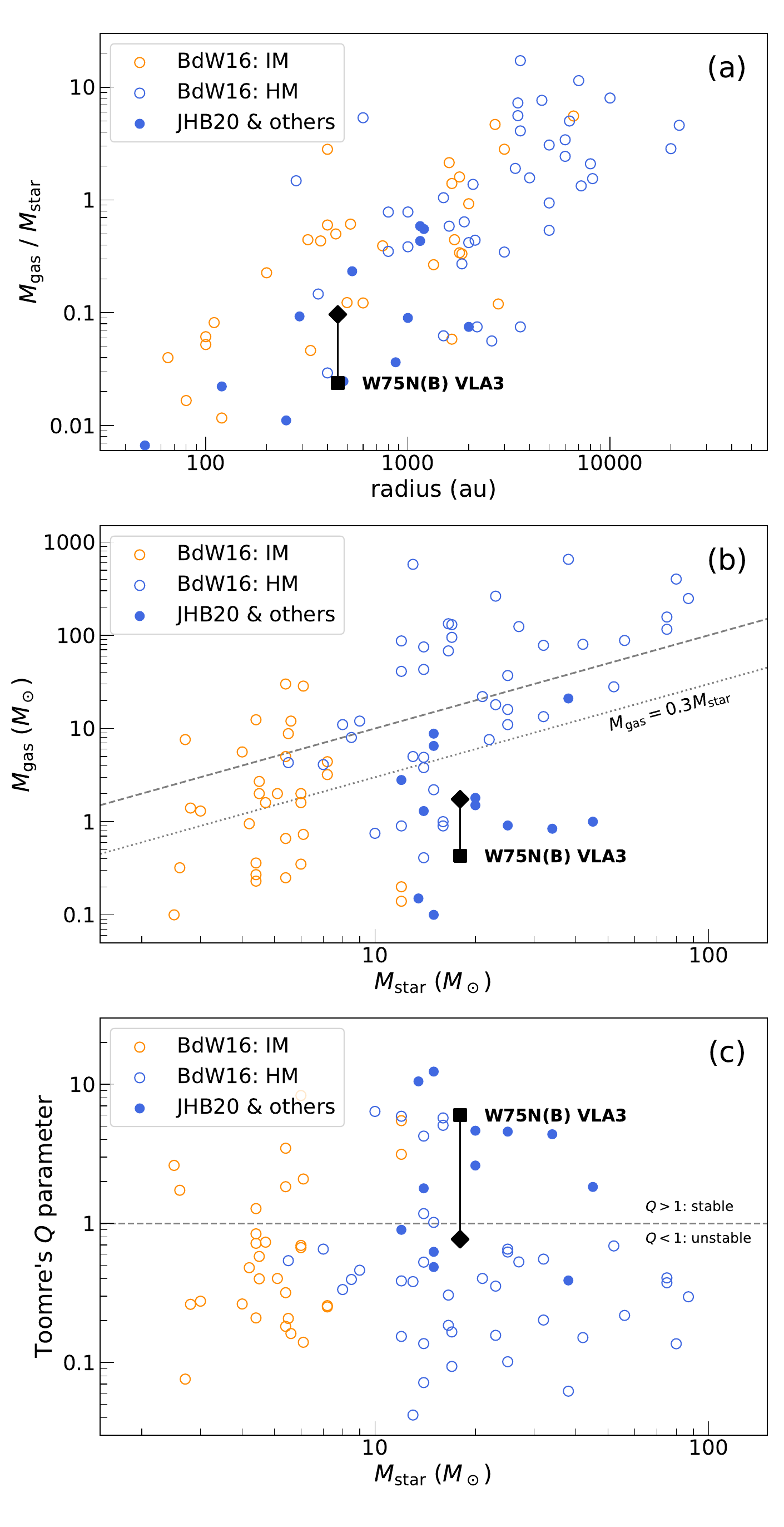}
\vspace{-0.5cm}
\caption{(\textit{panel~a}) Gas to star mass ratio, $M_\mathrm{gas}/M_\mathrm{star}$ as a function of the radius of the rotating structure. Blue and orange open circles depict rotating structures around intermediate-mass and high-mass YSOs, respectively, from \citet[][]{BeltrandeWit2016}. Filled blue circles correspond to high-angular resolution discs around massive stars \citep[see Table~7 of][and references therein]{Johnston2020}. The black symbols mark the location of the disc around the massive YSO W75N(B)-VLA3 for two different gas temperatures (50 and 200~K), corresponding to 1.74~$M_\odot$ (diamond symbol) and 0.43~$M_\odot$ (square symbol), respectively. (\textit{panel~b}) Gas mass $M_\mathrm{gas}$ of the rotating structure as a function of the central stellar mass $M_\mathrm{star}$. Blue and orange symbols as in panel~a. The dashed black line indicates $M_\mathrm{gas}=M_\mathrm{star}$ while the dotted black line indicates $M_\mathrm{gas}=0.3\,M_\mathrm{star}$, the maximum disc mass to allow for disc stability (see Sect.~\ref{sec:diskproperties}). The disc around VLA3 is located below this line suggesting it is a stable, rotating structure. (\textit{panel~c}) Toomre's $Q$ parameter as a function of stellar mass $M_\mathrm{star}$. The symbols are the same as in the previous panels. The black dashed line indicates $Q=1$ distinguishing gravitationally-unstable rotating structures ($Q<1$) and stable objects ($Q>1$).}
\label{fig:diskProperties_v1}
\end{figure}

\begin{figure}
\centering
\includegraphics[width=0.88\columnwidth]{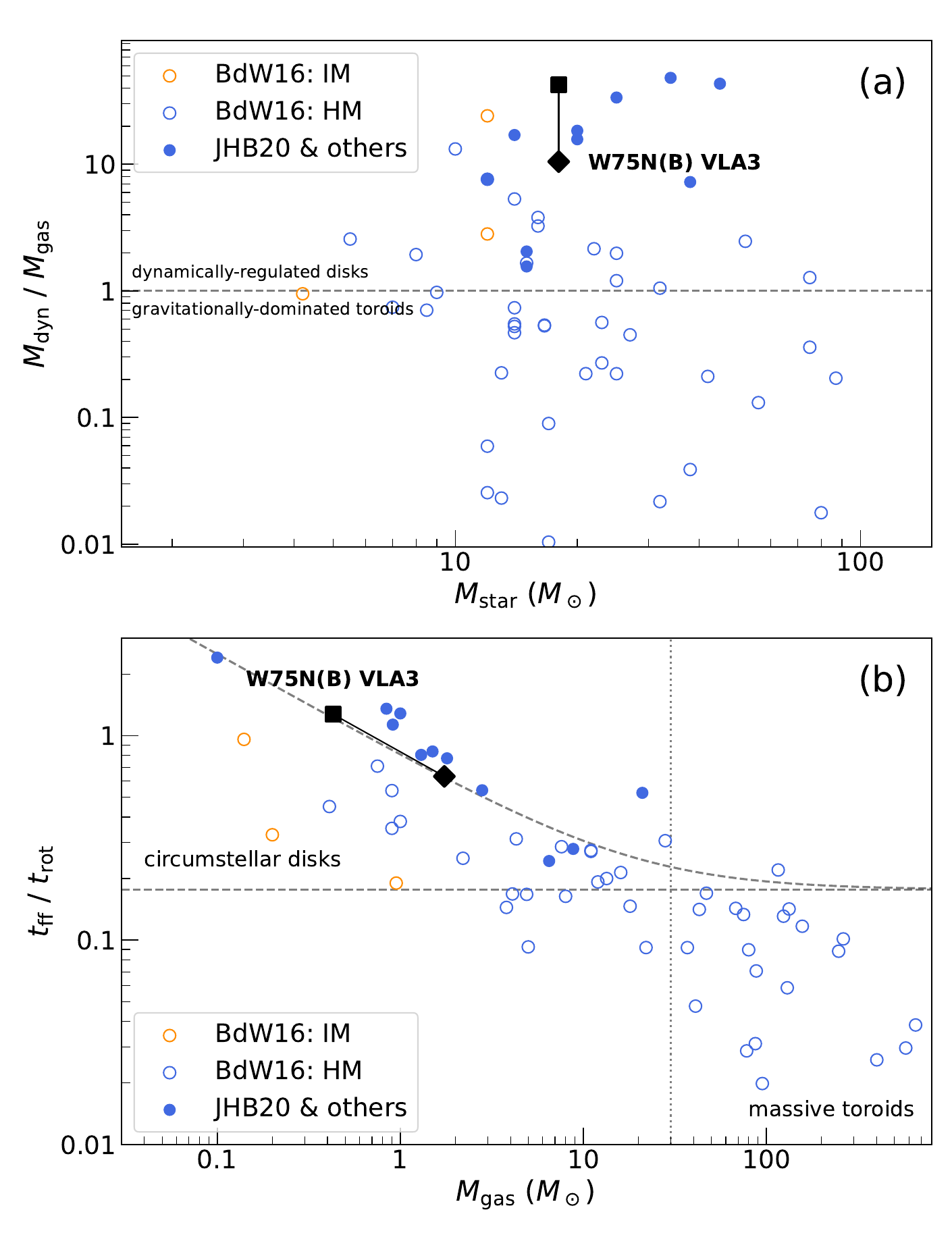}
\vspace{-0.5cm}
\caption{(\textit{panel~a}) Dynamical mass to gas mass ratio. $M_\mathrm{dyn}/M_\mathrm{gas}$, as a function of the central mass $M_*$ for rotating structures around high-mass (blue) and intermediate-mass (orange) YSOs. Blue and orange, open circles correspond to data collected in \citet[][]{BeltrandeWit2016}, while blue filled circles correspond to recent high-resolution studies \citep[see Table~7 of][and references therein]{Johnston2020}. The black symbols mark the position of the circumstellar disc around VLA3 for two different gas temperatures (50 and 200~K), corresponding to  1.74~$M_\odot$ (diamond symbol) and 0.43~$M_\odot$ (square symbol), respectively. The horizontal, dashed-black line indicates $M_\mathrm{dyn}=M_\mathrm{gas}$ and distinguishes those structures that are dynamically regulated (above the line) and gravitationally dominated (below the line). (\textit{panel~b}) Free-fall timescale to rotational period, $t_\mathrm{ff}/t_\mathrm{rot}$ as function of the gas mass $M_\mathrm{gas}$ of the rotating structure. Blue, orange and black symbols depict high-mass objects, intermediate-mass objects, and VLA3, respectively, as in panel~a. The dashed lines correspond to spheres of mass $M_\mathrm{gas}$ containing a star of mass $M_*$ at the center, in which the gas is rotationally supported against the gravity of both the gas plus the star. The horizontal and curved lines corresponds to $M_*=0$~$M_\odot$ and $M_*=10$~$M_\odot$, respectively. The vertical dotted line highlights two different populations: circumstellar discs to its left, and massive toroids to its right. See Sect.~\ref{sec:diskproperties} for details. Note that in both panels, there are less intermediate-mass objects (orange symbols) plotted compared to the panels in Fig.~\ref{fig:diskProperties_v1}. This is due to the lack of kinematic measurements for some of these objects that prevent us from deriving the dynamical mass and rotational period.}
\label{fig:diskProperties_v2}
\end{figure}

In Figs.~\ref{fig:diskProperties_v1} and \ref{fig:diskProperties_v2} we show the properties of the rotating structure found around W75N(B)-VLA3 alongside those of other potential rotating discs or toroids associated with intermediate and high-mass YSOs. For this, we use the review by \citet[][]{BeltrandeWit2016}, as well as recent observational results obtained at high-angular resolution \citep[see][and references therein]{Johnston2020, Ginsburg2023}. Generally, we refer to structures smaller than 1000~au as discs, and those larger than 5000~au as toroids (see also below). Note that the work by \citet{BeltrandeWit2016} primarily consists of data collected before the recent ALMA era of high-angular resolution studies. These earlier studies typically used low-energy transitions of dense gas tracers such as CH$_3$CN and CH$_3$OH to characterize the dense, rotating structures around massive YSOs. In contrast, the higher resolution and sensitivity observations of the new studies, enable the use of high-excited transitions of species such as H$_2$O, NaCl, and KCl \citep[e.g.,][]{Ginsburg2023}. For the comparison of VLA3 with other sources, we follow the approach used by \citet{BeltrandeWit2016} and use the gas mass contained in the rotating structure ($M_\mathrm{gas}$) and its radius, the central (proto)stellar mass ($M_\mathrm{star}$) derived from luminosity estimates, the dynamical mass ($M_\mathrm{dyn}$) derived from PV-plot analysis, and different timescales to evaluate the stability and dynamical status of the rotating object around VLA3.

In panel~(a) of Fig.~\ref{fig:diskProperties_v1}, we compare the gas-to-star mass ratio ($M_\mathrm{gas}/M_\mathrm{star}$) as a function of the radius of the rotating structure. The location of VLA3 in the diagram is in agreement with the correlation found between the distance-independent $M_\mathrm{gas}/M_\mathrm{star}$ and the radius of the circumstellar structures of other intermediate-mass (IM) and high-mass (HM) YSOs. As stated in Sect.~\ref{sec:continuum}, the mass of the rotating structure around VLA3 ($M_\mathrm{gas}\approx0.43$--1.74~$M_\odot$ based on the mm continuum emission) is about one tenth of the mass of the central object ($M_\mathrm{star}\approx17$--19~$M_\odot$ based on the relation between the radio and bolometric luminosities\footnote{We note that using the dynamically-derived mass of $\approx16\pm4$~$M_\odot$ for VLA3 (see Sect.~\ref{sec:pvplots}) does not significantly impact the results discussed here, since both estimates for the stellar masses are in good agreement.}). This $M_\mathrm{gas}/M_\mathrm{star}$ ratio estimated for VLA3 is similar to those recently found in other accretion discs around massive protosars \citep[see filled blue symbols in Fig.~\ref{fig:diskProperties_v1};][]{Johnston2020, Sanna2025}. Panel~(b) illustrates this mass relationship for VLA3 in comparison to the other rotating structures. The figure shows that the mass of the rotating structure (disc or toroid) is in general proportional to the mass of the central object. It is worth noting that in the earlier studies reported in \citet[][]{BeltrandeWit2016}, a significant number of rotating structures around IM and HM objects have $M_\mathrm{gas}>M_\mathrm{star}$, in particular for structures with radii generally $\gtrsim 300$~au. Consequently, these rotating structures are likely to be self-gravitating and are probably not in Keplerian rotation, as the gravitational potential of the system is primarily influenced by the surrounding circumstellar structure rather than the central (proto)stellar mass. Based on different studies \citep[e.g.][]{Shu1990, LaughlinBodenheimer1994, Cesaroni2007} discs can be stable if their mass is $\lesssim0.3 M_\mathrm{star}$. For larger disc masses (i.e., $>0.3 M_\mathrm{star}$), gravitational instabilities might develop and rapidly fragment the disc \citep[e.g.,][]{Rosen2019, Ahmadi2019}. This is expected to happen in the massive and rotating large structures found around massive YSOs, the so-called toroids \citep[see, e.g.,][]{Beltran2011}. These toroids, although not being stable entities, might host stable Keplerian-rotating discs in their interiors, which would be only accessible with high-angular resolution observations \citep[e.g.,][]{SanchezMonge2013a, Johnston2015, Cesaroni2017}. The existence of these stable structures is seen in the recent higher-angular resolution studies \citep[see e.g.,][]{Johnston2020, Ginsburg2023} that are able to resolve and trace the inner regions of the rotating structures (see filled blue circles in Fig.~\ref{fig:diskProperties_v1} panel~b). These observations show that in most cases these discs have masses less than 0.3 times that of the central (proto)star, as also found towards VLA3, suggesting that they are likely close to Keplerian rotation (with the mass of the central object being the dominant contributor). The high angular resolution observations of W75N(B) conducted with ALMA, have resolved the rotating disc in VLA3, adding it to the list of about a dozen such discs identified around massive protostars so far. 

The stability of these structures can be evaluated by estimating the Toomre's $Q$ parameter \citep[][see also \citealt{Ahmadi2023}]{Toomre1964}, which is defined as
\begin{equation}
Q = \frac{c_\mathrm{s}\kappa}{\pi G \Sigma}
\end{equation}
where $c_\mathrm{s}$ is the sound speed\footnote{The sound speed of the gas is calculated as $\sqrt{(k_\mathrm{B} T)/(\mu m_\mathrm{H})}$, where $k_\mathrm{B}$ is the Boltzmann constant, $m_\mathrm{H}$ is the hydrogen mass, and $\mu$ the mean mass assumed to be 2.3. For the temperature $T$ of the gas we use the values 50 and 200~K as in Sect.~\ref{sec:continuum}} of the gas, $\Sigma$ the surface density, and $\kappa$ the epicyclic frequency of the disc. For Keplerian rotation, the epicyclic frequency corresponds to the angular velocity $\Omega$ \citep[see][]{BeltrandeWit2016}, which following \citet{Cesaroni2007} can be estimated as $\sqrt{G M_\mathrm{total}/R^3}$, where $M_\mathrm{total}$ is the star plus disc total mass of the system and $R$ is the disc radius. Following \citet{BeltrandeWit2016}, the surface density is estimated as $\Sigma=M_\mathrm{gas}/(\pi R^2$). For a stellar mass of 16~$M_\odot$ and a disc radius of 450~au, we derive a Toomre's $Q$ parameter of 0.8 if the disc has a temperature of 50~K and a mass of 1.74~$M_\odot$, or a value of 6.2 if we consider a temperature of 200~K and a resulting disc mass of 0.43~$M_\odot$. These values are shown in the panel~(c) of Fig.~\ref{fig:diskProperties_v1} and compared to other discs and toroids. Based on this, VLA3 appears stable against gravitational instabilities, with a Toomre's $Q$ parameter comparable to that of the dozen discs resolved around massive YSOs so far. Panel~(c) of Fig.~\ref{fig:diskProperties_v1} shows how high-resolution observations are able to detect discs that populate the upper-right area of the diagram, which corresponds to stable discs rotating around massive stars ($M_\mathrm{star}>10$~$M_\odot$).

In Fig.~\ref{fig:diskProperties_v2}, we evaluate the dynamical status of the disc around VLA3. In panel~(a), we compare the dynamical mass to gas mass ratio of the discs/toroids against the central (proto)stellar mass. The dynamical mass ($M_\mathrm{dyn}$) is calculated assuming equilibrium between centrifugal and gravitational forces as $M_\mathrm{dyn} = v_\mathrm{rot}^2 R/G \sin^2{i}$, where $v_\mathrm{rot}$ is the rotational velocity at a radius $R$, which for VLA3 is estimated to be 6~km~s$^{-1}$ at 450~au, which results in a mass of $\approx16$~$M_\odot$ as derived from the Keplerian fits (see Sect.~\ref{sec:pvplots}). The dashed, horizontal line in the figure distinguishes two regimes: structures with $M_\mathrm{dyn}/M_\mathrm{gas}>1$ that are centrifugally supported, and objects with $M_\mathrm{dyn}/M_\mathrm{gas}<1$ that are gravitationally dominated. The high $M_\mathrm{dyn}/M_\mathrm{gas}$ value for VLA3 places this disc as a dynamically stable rotating structure comparable to some of the newest discs detected around massive YSOs.

Following \citet{Beltran2011} and \cite{SanchezMonge2014}, in Fig.~\ref{fig:diskProperties_v2} panel~(b), we plot the ratio of the free-fall timescale ($t_\mathrm{ff}$) to the rotational period ($t_\mathrm{rot}$) against $M_\mathrm{gas}$ of the rotating structure. The free-fall timescale is proportional to the dynamical timescale needed to refresh the material of the rotating structure, while the rotational period at the outer radius is the time needed by the rotating structure to stabilise after incorporating new accreted material. Note that $t_\mathrm{ff}$ could be slightly overestimated since spherical symmetry is assumed \citep[see also][]{BeltrandeWit2016}. The dashed curves in the figure correspond to the theoretical $t_\mathrm{ff}/t_\mathrm{rot}$ curves for spherical clouds of mass $M_\mathrm{gas}$ containing a star of mass $M_*$ at the center, in which the gas is rotationally supported against the gravity of the gas plus the star. These curves are expressed as $t_\mathrm{ff}/t_\mathrm{rot}=[(M_\mathrm{gas}+M_*)/(32\,M_\mathrm{gas})]^{1/2}$ \citep[see][]{Beltran2011, SanchezMonge2014}. The results shown in Fig.~\ref{fig:diskProperties_v2} confirm that the rotating structure around VLA3 is in agreement with other centrifugally-regulated circumstellar discs, and distinguishable from gravitationally-unstable massive toroids.

Overall, the high-angular resolution study of the disc in W75N(B)-VLA3 contributes to the growing number of recently identified discs surrounding massive YSOs. It also suggests that highly excited transitions of molecules such as SO$_2$, H$_2$O, or NaCl might be more effective than the traditionally used transitions of complex organic molecules like CH$_3$CN and CH$_3$OH for resolving the inner regions of gravitationally-stable, Keplerian-like rotating discs.

%
\subsection{Origin of the RRL emission in W75N(B)-VLA3}\label{sec:RRLs}

In Sect.~\ref{sec:ionized-gas}, we have reported the detection of the RRL H30$\alpha$ emission towards VLA3. The limited spectral resolution, approximately 20~km~s$^{-1}$, hampers our ability to fully characterise the shape and velocity structure of the RRL; however, the spatial relationship with both the bright 1.3~mm continuum source and the radio continuum source reported by \citet{CarrascoGonzalez2010a} suggests that the H30$\alpha$ emission could be associated with the VLA3 thermal radio jet. Specifically, according to \citet{Anglada2018}, the ratio of the RRL and continuum flux densities, $S_\mathrm{L}/S_\mathrm{C}$, for a biconical thermal jet with constant velocity and ionised fraction is given by
\begin{equation}\label{eq:RRLs}
\frac{S_\mathrm{L}}{S_\mathrm{C}} =
\left[
0.25
\left(\frac{\nu_\mathrm{L}}{\mathrm{GHz}}\right)^{1.1}
\left(\frac{T_\mathrm{e}}{10^4~\mathrm{K}}\right)^{-1.1}
\left(\frac{\Delta v}{\mathrm{km~s}^{-1}}\right)^{-1}
+1\right]^{2/3}-1
\end{equation}
after considering an ionised helium to ionised hydrogen ratio of 0.1, with $\nu_\mathrm{L}$ the frequency of the RRL, $T_\mathrm{e}$ the electron temperature of the ionised gas, and $\Delta v$ the full width of the RRL. The radio continuum emission at 2~cm has a flux density of 5.7~mJy \citep[see][]{CarrascoGonzalez2010a} which is extrapolated to a continuum flux density of 25~mJy at 1.3~mm, assuming a spectral index of $+0.6$ (see also Sect.~\ref{sec:continuum}). Using Eq.~(\ref{eq:RRLs}), one expects an H30$\alpha$ thermal flux density of $\approx37$~mJy, assuming a temperature of 10\,000~K and a line width of 100~km~s$^{-1}$. This expected H30$\alpha$ flux density is in good agreement with the observed line intensity (see Sect.~\ref{sec:ionized-gas}), after considering that about 15\% of the observed flux comes from a blended $^{33}$SO$_2$ line. Removing this contamination from the measured flux results in an observed H30$\alpha$ flux density of $\approx36$~mJy, consistent with the theoretical expectation.

The observed velocity of the H30$\alpha$ line, which is close to the ambient velocity (see Sect.~\ref{sec:ionized-gas}), could also be consistent with a jet moving almost on the plane of the sky. In addition, the observed RRL line width of $\simeq100$~km~s$^{-1}$, might be explained in an approximate way by considering a conical jet with an aperture semi-angle $\theta_\mathrm{j}$, following Rivera-Ort\'iz, Lizano and Cant\'o (2025; in preparation). For this, we assume that the flow within the jet is radial with respect to the powering source, and has a constant and uniform velocity $\mathrm{V}_\mathrm{j}$. We consider an observer's line of sight that makes an angle $\theta_\mathrm{o}$ with the jet's symmetry axis. Based on these assumptions, an emission line originating from the jet will be confined to the line-of-sight velocity interval $[\mathrm{V}_\mathrm{min}, \mathrm{V}_\mathrm{max}]$, centered at $\mathrm{V}_\mathrm{c}$ = $(\mathrm{V}_\mathrm{min} + \mathrm{V}_\mathrm{max})/2$. The minimum and maximum velocities are given by
\begin{equation}
\mathrm{V}_\mathrm{min} = \mathrm{V}_\mathrm{j} \cos(\theta_\mathrm{o} + \theta_\mathrm{j}), ~\mathrm{and} ~\mathrm{V}_\mathrm{max} = \mathrm{V}_\mathrm{j} \cos(\theta_\mathrm{o} - \theta_\mathrm{j}),
\end{equation}
respectively. The expression for $\mathrm{V}_\mathrm{max}$ is valid for $\theta_\mathrm{o}>\theta_\mathrm{j}$. For $\theta_\mathrm{o}<\theta_\mathrm{j}$, we have $\mathrm{V}_\mathrm{max}=\mathrm{V}_\mathrm{j}$. Therefore, the emission line will have a full width at zero intensity of
\begin{equation}
\Delta \mathrm{V} = \mathrm{V}_\mathrm{max} - \mathrm{V}_\mathrm{min} = 2\mathrm{V}_\mathrm{j}\sin(\theta_\mathrm{o})\sin(\theta_\mathrm{j}), 
\end{equation}
centered at
\begin{equation}
\mathrm{V}_\mathrm{c} = \mathrm{V}_\mathrm{j}\cos(\theta_\mathrm{o})\cos(\theta_\mathrm{j}).
\end{equation}
A counter-jet (i.e., a jet with identical physical conditions but moving in an opposite direction) will produce a `mirror' emission line in the velocity interval $[-\mathrm{V}_\mathrm{min}, -\mathrm{V}_\mathrm{max}]$, centered at $\mathrm{V}_\mathrm{c} = -\mathrm{V}_\mathrm{j}\cos(\theta_\mathrm{o})\cos(\theta_\mathrm{j})$, with the same full width as the jet's line. If the jet is unresolved and its axis is close to the plane of the sky (i.e., $\theta_\mathrm{o} \simeq \pi/2$), the two lines (those from the jet and counter jet) will blend in a single line, centered at zero velocity (the velocity of the system) and having a full width at zero intensity of
\begin{equation}
\begin{split}
\Delta \mathrm{V} = 2\mathrm{V}_\mathrm{max}
    & = 2\mathrm{V}_\mathrm{j}\cos(\theta_\mathrm{o} - \theta_\mathrm{j}) \\
    & \simeq 2\mathrm{V}_\mathrm{j}\sin(\theta_\mathrm{j}), ~\mathrm{for} ~\theta_\mathrm{o} \simeq \pi/2.
\end{split}
\end{equation} 

Applying this scenario to our observations, if we take a velocity for the jet flow of $\mathrm{V}_\mathrm{j}=500$~km~s$^{-1}$ \citep[similar to those observed in other massive protostars, e.g.,][]{Curiel2006, Anglada2018}, and a line width of $100$~km~s$^{-1}$ (or $300$~km~s$^{-1}$, based on the high-velocity emission seen in Fig.~\ref{fig:H30a_channels}), we obtain $\theta_\mathrm{j} \simeq 6^{\circ}$ (or $\simeq 17^{\circ}$ after considering the highest observed velocities), which is consistent with the aperture semi-angle derived from the size of the radio jet at a wavelength of 3.6~cm \citep[$\simeq 9^{\circ}$;][]{CarrascoGonzalez2010a, RodriguezKamenetzky2020}. Similarly, considering an inclination angle $\theta_\mathrm{o} \approx 50^\circ$ (see Sect.~\ref{sec:pvplots}) and the aperture semi-angle $\theta_\mathrm{j}\simeq9^\circ$ measured for the VLA3 jet \citep[][]{CarrascoGonzalez2010a}, a jet flow of $\mathrm{V}_\mathrm{j}\approx 70$--$200$~km~s$^{-1}$ would suffice to reproduce the observed H30$\alpha$ line width of $100$ and $300$~km~s$^{-1}$. A more detailed model, able to draw more robust conclusions, would require H30$\alpha$ line data with higher spectral and angular resolution than our current data provide.

All this suggests that W75N(B)-VLA3 is likely the first radio jet in which a thermal RRL has been detected. Nevertheless, we also considered alternative scenarios. For example, the observed RRL emission could originate from an unresolved, very young ultracompact or hypercompact H{\sc ii} region (UC/HC H{\sc ii}) \citep[e.g.,][]{Sewilo2004, Keto2008}. This scenario, however, seems unlikey given the extremely broad line width of $\sim$$100$~km~s$^{-1}$ observed for the H30$\alpha$ line. In this case, the expected static (non-expanding) H{\sc ii} region thermal width of $\sim20$~km~s$^{-1}$ cannot alone explain the observed line width, neither can the typical H{\sc ii} region expansion velocities of $\sim10$--$15$~km~s$^{-1}$ \citep[e.g.,][]{Faerber2025}. Moreover, although extremely dense, hypercompact H{\sc ii} regions may have broader lines \citep[e.g.,][]{Keto2008, RiveraSoto2020}, they only reach widths of $\approx50$~km~s$^{-1}$ at cm wavelengths, and $\approx30$~km~s$^{-1}$ at mm wavelengths. Another possible scenario is that the RRL emission arises from an inner ionised disc \citep[see, e.g., the case of G17.64$+$0.16 by][see also \citealt{Tanaka2016, JaquezDominguez2025}]{Maud2019}. However, the tentative north-south spatial shift and velocity gradient reported in Fig.~\ref{fig:H30a_positions} for the H30$\alpha$ line, if confirmed with further studies, would favour the radio jet scenario over the ionised disc. Moreover, based on the Keplerian-like profiles shown in Fig.~\ref{fig:PVcut}, one would expect full widths of $\sim50$~km~s$^{-1}$ for RRLs tracing the ionised disc at our current spatial resolution. This is about a factor two to six lower than the observed H30$\alpha$ widths ($\sim100$~km~s$^{-1}$, and $\sim300$~km~s$^{-1}$ at zero intensity). All this leaves the jet scenario as the most plausible explanation for the origin of the H30$\alpha$ emission in W75N(B)-VLA3. Should this be confirmed with new observations, VLA3 could represent the first protostellar jet with detected thermal radio recombination lines.
 
Robust detections of thermal RRLs associated with protostellar thermal radio jets have, to date, remained elusive. This is largely due to the limited sensitivity of current centimetre-wavelength interferometers. While recombination maser lines have been reported towards some disc-outflow systems (e.g., MWC349A: \citealt{MartinPintado1989}; MonR2-IRS2: \citealt{JimenezSerra2013}; G45.47+0.05: \citealt{Zhang2019}) including the Cepheus A HW2 protostellar jet \citep[][]{JimenezSerra2011}, general thermal RRL detections are a significant observational challenge. Detecting multiple thermal RRLs arising within the radio jets at both centimetre (e.g., with the ngVLA, and/or SKA) and millimetre (e.g., with ALMA, SMA, and/or NOEMA) wavelengths will constitute a profound breakthrough. Such detections would, for instance, enable us to analyse the morphology and 3D kinematics of protostellar ionised radio jets by examining their line-of-sight velocities and proper motions, thus providing fundamental information to understand their collimation mechanisms \citep[see, e.g.,][and references therein]{Anglada2018, Tanaka2016, JaquezDominguez2025}. Our findings identify W75N(B)-VLA3 as a unique candidate for undertaking this kind of studies.

%
\section{Conclusions}\label{sec:conclusions}

This work presents a study of the high-mass protostar W75N(B)-VLA3 using ALMA data at high angular resolution ($\approx0\farcs12$, corresponding to $\approx156$~au). We have used the continuum emission at 1.3~mm, together with spectral line emission from different species (e.g., SiO, H$_2$CO, HC$_3$N, SO$_2$, H$_2$O, H30$\alpha$) to characterize the morphology and properties of the dense gas around the massive protostar. Our principal findings include:

\begin{findings}

\item[--] {\it Keplerian-like disc around VLA3}: ALMA continuum and molecular line observations (specifically of SO$_2$, H$_2$CO, HC$_3$N, HNCO, CH$_3$CN, and H$_2$O) reveal a compact, rotating molecular disc with a radius of $\sim$450~au surrounding VLA3. The disc has an estimated gas and dust mass of $\approx0.43$--1.74~$M_\odot$ and its velocity profile is consistent with Keplerian motions around a protostar with mass $\approx16$~$M_\odot$. This dynamical mass aligns well with the stellar mass of $\approx17$--$19$~M$_\odot$ derived from the correlation between radio continuum and bolometric luminosities. The low disc-to-star mass ratio (approximately one-tenth of the central object's mass) indicates that the disc is a stable, centrifugally supported structure, rather than a gravitationally unstable toroid. Calculations of the Toomre's Q parameter further support its stability against gravitational instabilities compared to other massive structures. The properties of the disc in VLA3 are comparable to those recently measured towards discs around massive YSOs when observed at high resolution and sensitivity. The disc is oriented perpendicular to the VLA3 thermal radio jet previously reported at cm wavelengths.

\vspace{0.2cm}
\item[--] {\it Disc asymmetry}: The disc around VLA3 exhibits two intensity peaks in the dense gas molecular tracers, with the southwestern peak appearing brighter. This difference in brightness suggests an asymmetrical disc structure, potentially caused by anisotropic accretion streams, a phenomenon observed in other young stellar objects. An elongated, slightly curved structure visible to the north-east in some dense gas tracers (e.g., HC$_3$N, H$_2$CO) might be indicative of such an accretion streamer. While the two-peak morphology might suggest a molecular cavity around the central source, it is more plausibly interpreted as an apparent effect due to high dust opacity in the optically thick central region. The vibrationally excited H$_2$O emission, which peaks close to the central source, may arise from the innermost, warmest regions of the disc, where dust would have sublimated, with some potential contribution from a north-south outflow.

\vspace{0.2cm}
\item[--] {\it Expanding molecular gas}: Blue-shifted absorption features observed in the SiO and H$_2$CO spectral lines, reminiscent of classical P-Cygni profiles, suggest the presence of expanding molecular gas in the near vicinity of VLA3. This expansion motions might originate from the outflow activity of the central source or an expanding photoionised region.

\vspace{0.2cm}
\item[--] {\it Detection of H30$\alpha$ emission towards the radio jet}: The H30$\alpha$ recombination line is spatially coincident with the thermal radio jet, raising the possibility that W75N(B)-VLA3 could be the first known protostellar radio jet from which thermal radio recombination lines have been detected. The extremely broad H30$\alpha$ full width of $\approx100$~km~s$^{-1}$ (or $\approx300$~km~s$^{-1}$ at zero intensity), indicative of high velocity motions likely associated with outflows and jets, together with a tentative H30$\alpha$ spatial shift and velocity gradient in the jet's direction, favour the radio jet origin as the most plausible scenario for this line. New observations at millimetre wavelengths with current instruments such as ALMA, NOEMA, and SMA, together with observations at centimetre wavelengths with ngVLA and SKA, will enable a detailed study and modelling of the RRLs emission and their origin in VLA3.

\end{findings}

In summary, these results establish W75N(B)-VLA3 as one of the few well-documented massive protostar-disc-jet systems with a disc radius $\lesssim$500~au, confirming its significance as a key system that offers critical observational constraints for high-mass star formation theories; and identify this source as a unique radio jet object, suitable for conducting an in-depth study of its RRL emission.

%
\section*{Acknowledgements}

We thank the anonymous referee for their constructive comments that have helped to improve the content and clarity of the manuscript.
A.S-M., J.M.T, and J.M.G.\ acknowledge support from the PID2023-146675NB grant funded by MCIN/AEI/10.13039/501100011033, and by the programme Unidad de Excelencia Mar\'{\i}a de Maeztu CEX2020-001058-M. A.S-M.\ also acknowledges support from the RyC2021-032892-I grant funded by MCIN/AEI/10.13039/501100011033 and by the European Union `Next GenerationEU'/PRTR.
J.F.G., G.A., and G.A.F.\ acknowledge support through the Severo Ochoa grant CEX2021-001131-S and PID2023-146295NB-I00PID, funded by MCIN/AEI/10.13039/5011000110. G.A.F.\ also gratefully acknowledges the Deutsche Forschungsgemeinschaft  (DFG) for funding through SFB\,1601 'Habitats of massive stars across cosmic time' (sub-project B1), and the University of Cologne and its Global Faculty programme. 
S.C.\ acknowledges the support of DGAPA PAPIIT grants IN104521 and IN107324 as well as CONAHCyT grant CF-2023-I-232.
C.G.\ acknowledges financial support by
the European Union NextGenerationEU RRF M4C2 1.1 project n. 2022YAPMJH; FAPESP (Funda\c{c}\~ao de Amparo \'a Pesquisa do Estado de S\~ao Paulo) under grant 2021/01183-8.
C.C.G.\ acknowledges support from UNAM DGAPA PAPIIT grant IG101224 and from CONAHCyT Ciencia de Frontera project ID 86372.
A.R.K.\ acknowledges support from Consejo Nacional de Investigaciones Científicas y Técnicas (CONICET), and support from the European Research Executive Agency HORIZON-MSCA-2021-SE-01 Research and Innovation programme under the Marie Sk{\l}odowska-Curie grant agreement number 101086388 (LACEGAL).
J.C.\ acknowledges support from the IG101125 (DGAPA) grant.
This paper makes use of ALMA data ADS/JAO.ALMA\#2019.1.00059.S. ALMA is a partnership of ESO (representing its member states), NSF (USA) and NINS (Japan), together with NRC (Canada), MOST and ASIAA (Taiwan), and KASI (Republic of Korea), in cooperation with the Republic of Chile. The Joint ALMA Observatory is operated by ESO, AUI/NRAO and NAOJ.
The data processing for this paper has been carried out using the Spanish Prototype of an SRC (SPSRC) service and support, funded by the Spanish Ministry of Science, Innovation and Universities, by the Regional Government of Andalusia, by the European Regional Development Funds and by the European Union NextGenerationEU/PRTR. 

\section*{Data Availability}

The ALMA data used in this work are available at the ALMA Science Archive under project number ADS/JAO.ALMA\#2019.1.00059.S



\bibliographystyle{mnras}
\bibliography{bibliow75} 




\appendix

%
\section{Additional figures}\label{app:additional}

\begin{figure*}
\centering
\includegraphics[width=1.0\textwidth]{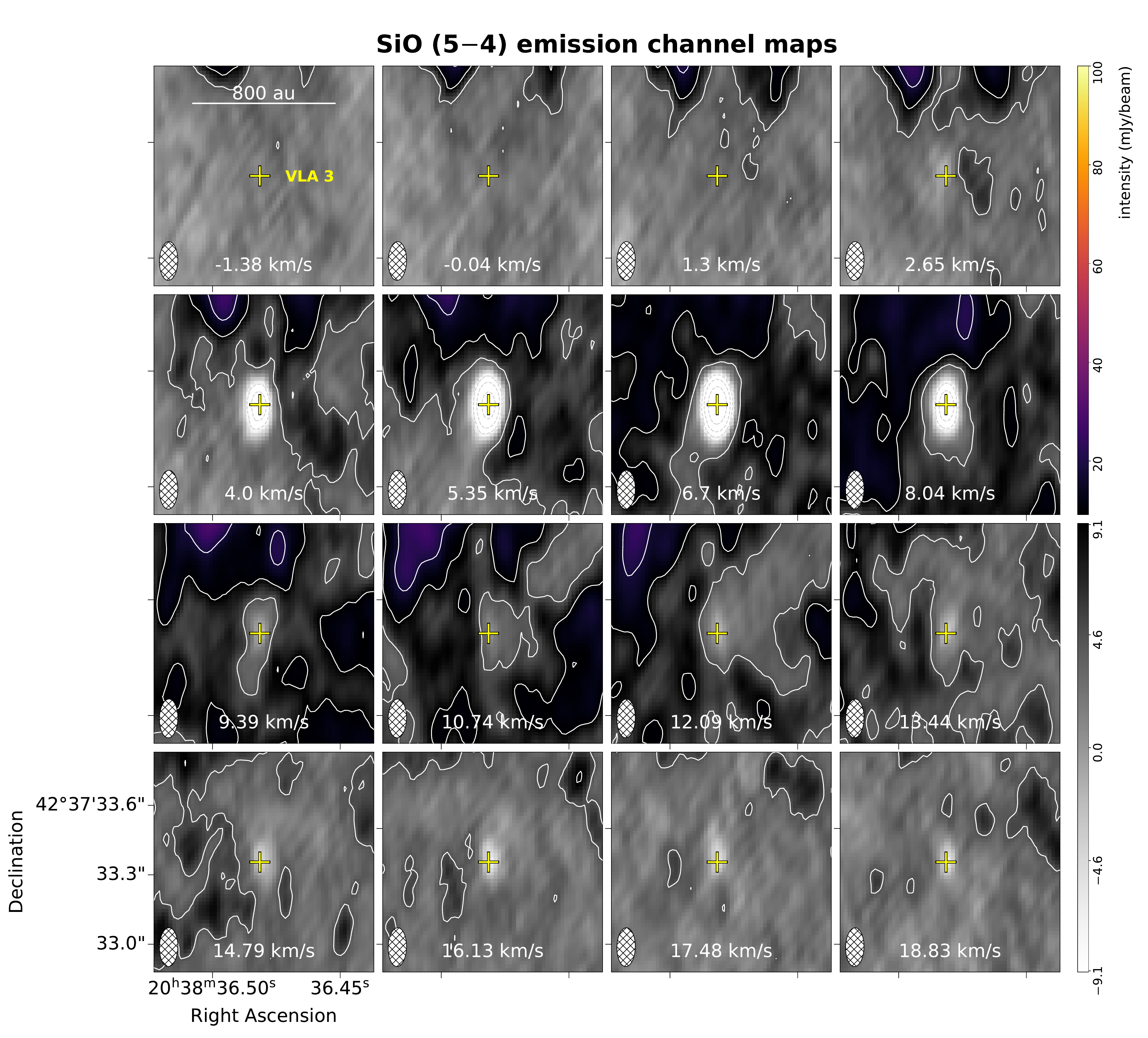}
\vspace{-0.5cm}
\caption{ALMA SiO\,(5--4) velocity channel maps. From top-left to bottom-right the velocity of each channel, indicated at the bottom of each panel, increases from $-1.38$~km~s$^{-1}$ to $+18.83$~km~s$^{-1}$, in steps of $\approx1.35$~km~s$^{-1}$. The dashed and solid contours mark the SiO emission at the levels $-32$, $-16$, $-8$, $-4$, 4, 8, 16, and 32 times the rms of the image ($\mathrm{rms}=1.14$~mJy~beam$^{-1}$ in channels of $\approx1.35$~km~s$^{-1}$ width; beam $=0\farcs167\times0\farcs079$, PA$=-1\fdg3$, depicted with a hatched ellipse in each panel). The yellow cross marks the position of the massive protostar VLA3. The SiO is mainly seen in absorption against the bright continuum emission of VLA3. This is particularly clear towards velocities V$_{\rm LSR}$ $\simeq$ 4 to 8~km~s$^{-1}$.}
\label{fig:SiO_channels}
\end{figure*}

\begin{figure*}
\centering
\includegraphics[width=1.0\textwidth]{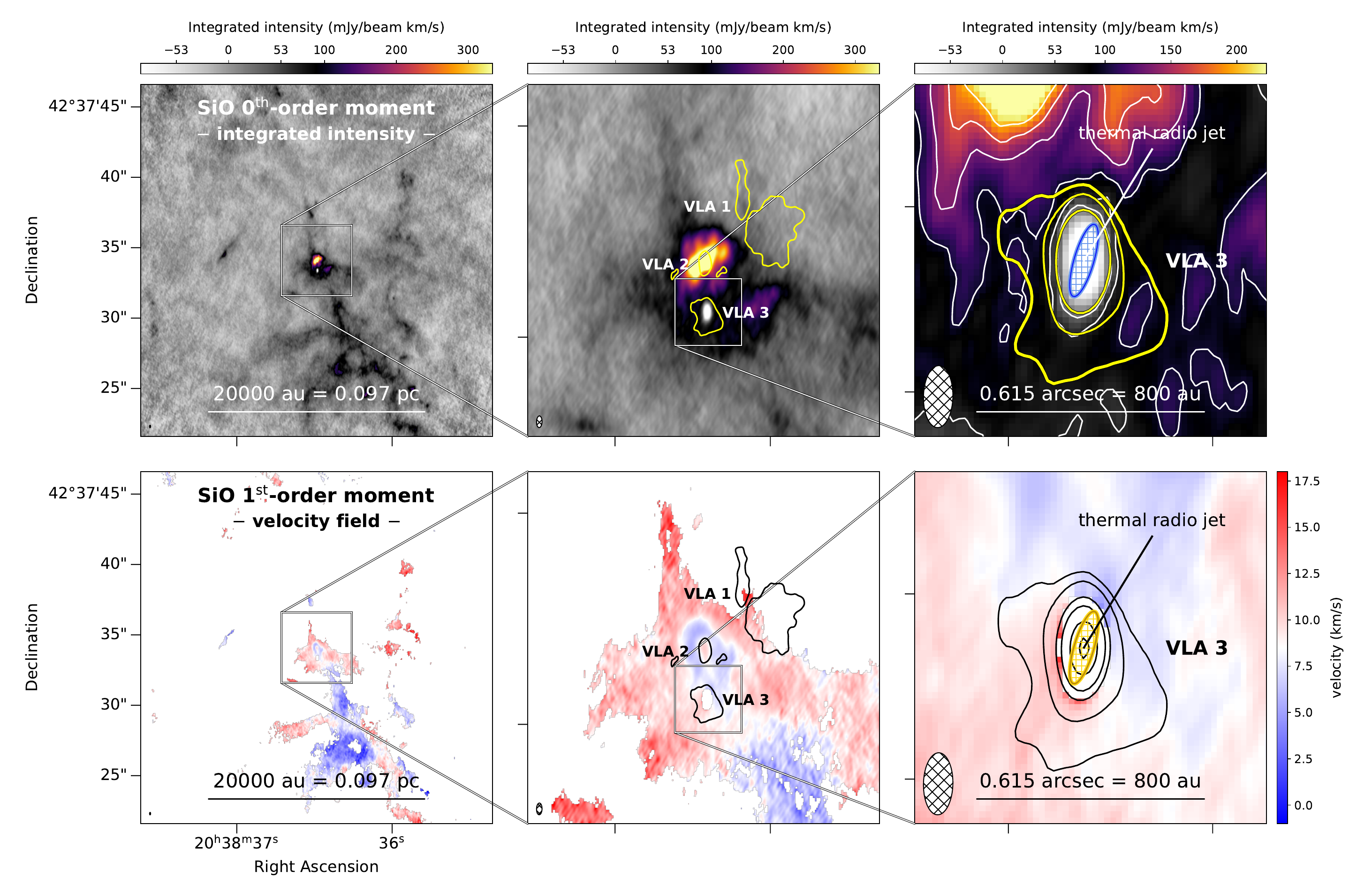}
\vspace{-0.5cm}
\caption{SiO\,(5--4) moment maps. The top and bottom rows show, respectively, the integrated intensity (0$^\mathrm{th}$ order) and the velocity field (1$^\mathrm{st}$ order). The velocity interval used to generate the moment maps ranges from $-1.38$ up to $+18.83$~km~s$^{-1}$ (see Fig.~\ref{fig:SiO_channels}). Columns from left to right show different zoomed-in views of the W75N(B) region, as described in Fig.~\ref{fig:continuum}. The ALMA 1.3~mm continuum emission (see Fig.~\ref{fig:continuum}) is shown in yellow contours in the top-row panels, and in black contours in the bottom-row panels. White contours in the top-right panel shows the SiO integrated emission at levels 3, 6, 9, and 12 times 11~mJy~beam$^{-1}$~km~s$^{-1}$, with the beam depicted by hatched ellipses. While strong SiO emission is observed towards VLA2 \citep[see][]{Gomez2023}, SiO is predominantly observed in absorption towards VLA3 (see also Fig.~\ref{fig:SiO_channels}). The blue and yellow ellipse in the right panels depicts the size and orientation of the thermal radio jet reported by \citet{CarrascoGonzalez2010a}.}
\label{fig:SiO_moments}
\end{figure*}

\begin{figure*}
\centering
\includegraphics[width=1.0\textwidth]{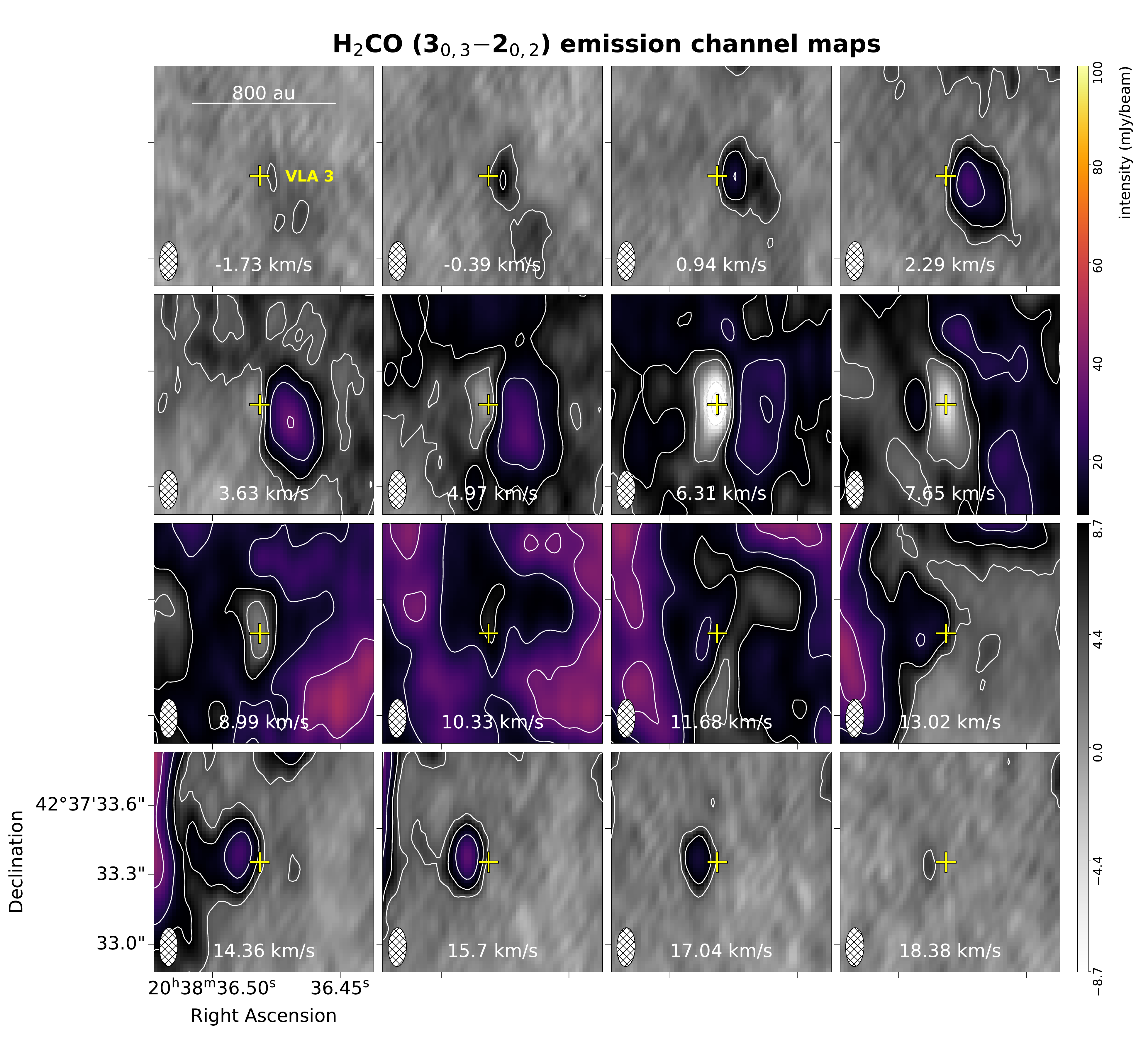}
\vspace{-0.5cm}
\caption{ALMA H$_2$CO\,(3$_{0,3}$--2$_{0,2}$) emission channel maps. From top-left to bottom-right the velocity of each channel, indicated at the bottom of each panel, increases from $-1.73$~km~s$^{-1}$ to $+18.38$~km~s$^{-1}$, in steps of $\approx1.35$~km~s$^{-1}$. The dashed and solid contours mark the H$_2$CO emission at the levels $-16$, $-8$, $-4$, 4, 8, 16, and 32 times the rms of the image ($\mathrm{rms}=1.09$~mJy~beam$^{-1}$ in channels of $\approx1.35$~km~s$^{-1}$ width; beam $=0\farcs168\times0\farcs079$, PA$=-1\fdg4$, depicted with a hatched ellipse in each panel). The yellow cross marks the position of the VLA3 YSO. The H$_2$CO emission shifts from southwest with respect to VLA3 at blue-shifted velocities to northeast at red-shifted velocities (with the systemic velocity at V$_\mathrm{LSR}\simeq$~8.9~km~s$^{-1}$, see Sect.~\ref{sec:discussion}). H$_2$CO is seen in absorption at the velocity of $\sim$6.3~km~s$^{-1}$.}
\label{fig:H2CO_channels}
\end{figure*}

\begin{figure*}
\centering
\includegraphics[width=1.0\textwidth]{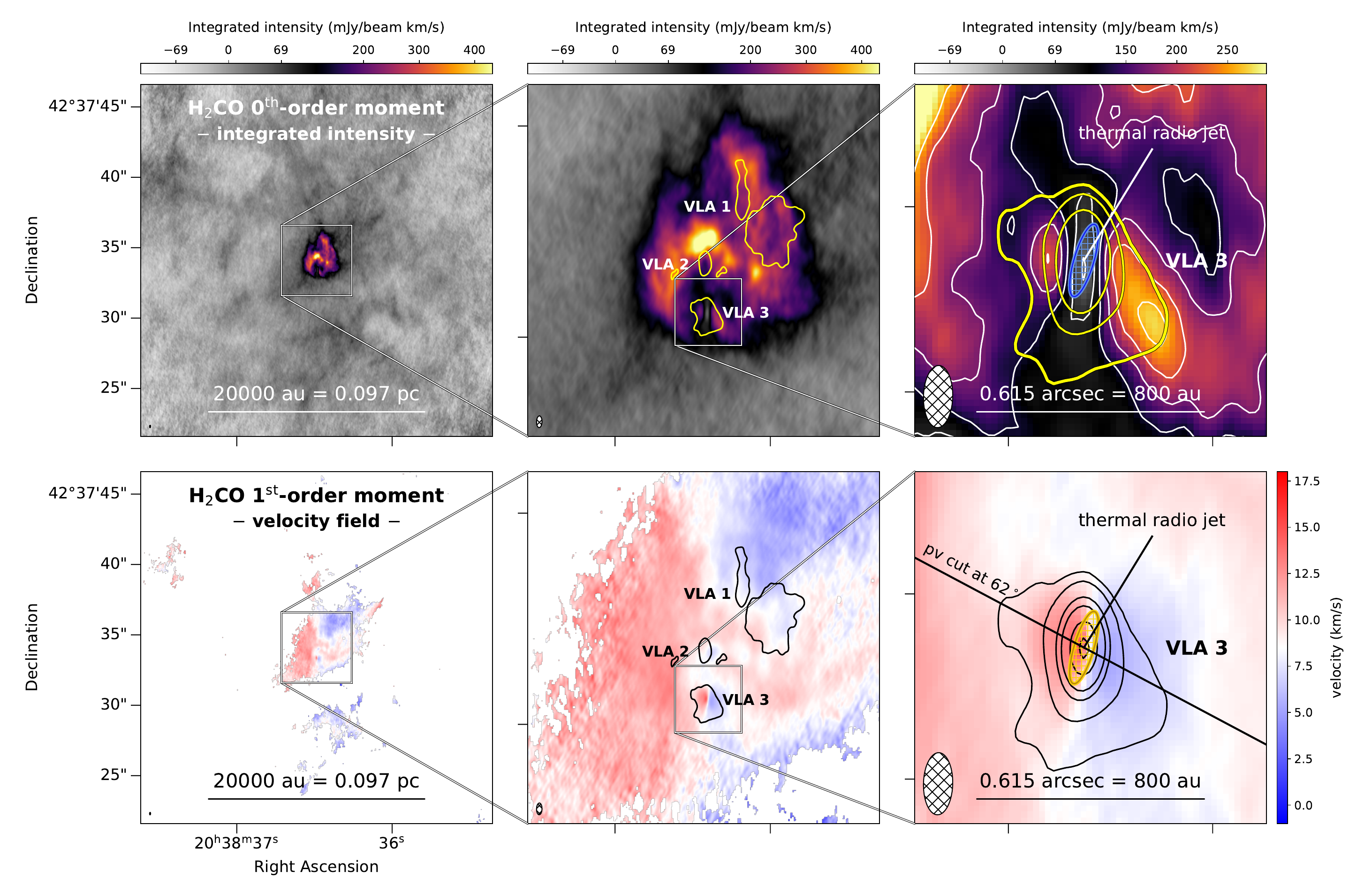}
\vspace{-0.5cm}
\caption{H$_2$CO\,(3$_{0,3}$--2$_{0,2}$) moment maps. Rows from top to bottom show: integrated intensity (0$^\mathrm{th}$ order); and velocity field (1$^\mathrm{st}$ order). The velocity interval used to generate the moment maps ranges from $-1.73$ up to $+18.38$~km~s$^{-1}$ (see Fig.~\ref{fig:H2CO_channels}). Columns from left to right show different zoomed-in views of the W75N(B) region, as described in Fig.~\ref{fig:continuum}. The ALMA 1.3~mm continuum emission (see Fig.~\ref{fig:continuum}) is shown in yellow contours in the top-row panels, and in black contours in the bottom-row panels. White contours in top-right panel shows the H$_2$CO integrated emission at levels 3, 6, 9, 12, 15, and 18 times 14~mJy~beam$^{-1}$~km~s$^{-1}$, with the beam depicted by hatched ellipses. The blue and yellow ellipse in the right panels depicts the size and orientation of the thermal radio jet reported by \citet{CarrascoGonzalez2010a}. The diagonal solid black line in the right panel of second row shows the direction of the position-velocity (PV) cut, with PA$=62^\circ$, shown in Fig.~\ref{fig:PVcut}.}
\label{fig:H2CO_moments}
\end{figure*}

\begin{figure*}
\centering
\includegraphics[width=1.0\textwidth]{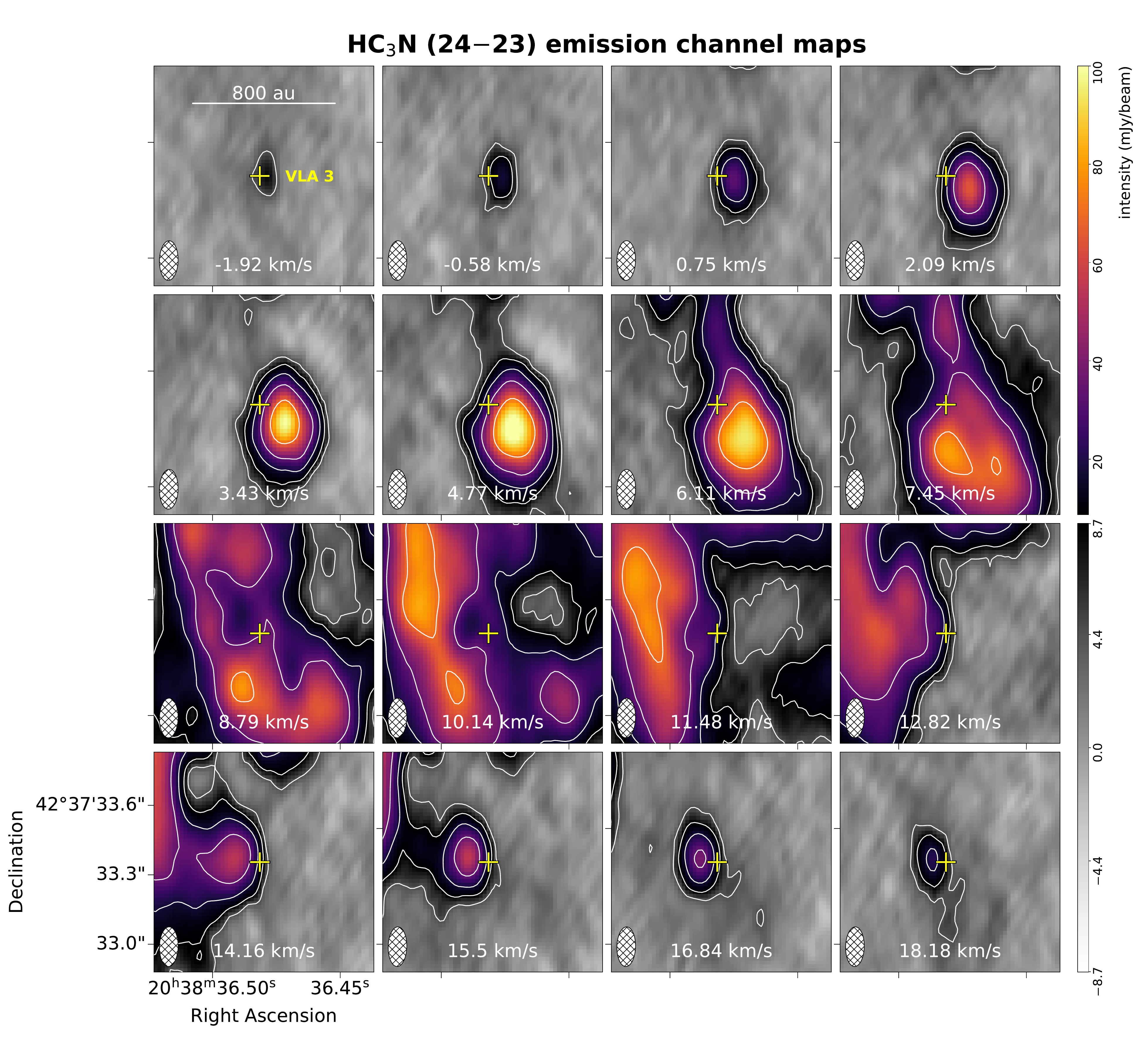}
\vspace{-0.5cm}
\caption{ALMA HC$_3$N\,(24--23) emission channel maps. From top-left to bottom-right the velocity of each channel, indicated at the bottom of each panel, increases from $-1.92$~km~s$^{-1}$ to $+18.18$~km~s$^{-1}$, in steps of $\approx1.35$~km~s$^{-1}$. The solid contours mark the HC$_3$N emission at the levels 4, 8, 16, 32, and 64 times the rms of the image ($\mathrm{rms}=1.09$~mJy~beam$^{-1}$ in channels of $\approx1.35$~km~s$^{-1}$ width; beam $=0\farcs171\times0\farcs081$, PA$=-1\fdg4$, depicted with a hatched ellipse in each panel). The yellow cross marks the position of VLA3. The HC$_3$N emission shifts from southwest with respect to VLA3 at blue-shifted velocities to northeast at red-shifted velocities (with the systemic velocity at V$_\mathrm{LSR}\simeq$~8.9~km~s$^{-1}$, see Sect.~\ref{sec:discussion}).}
\label{fig:HC3N_channels}
\end{figure*}

\begin{figure*}
\centering
\includegraphics[width=1.0\textwidth]{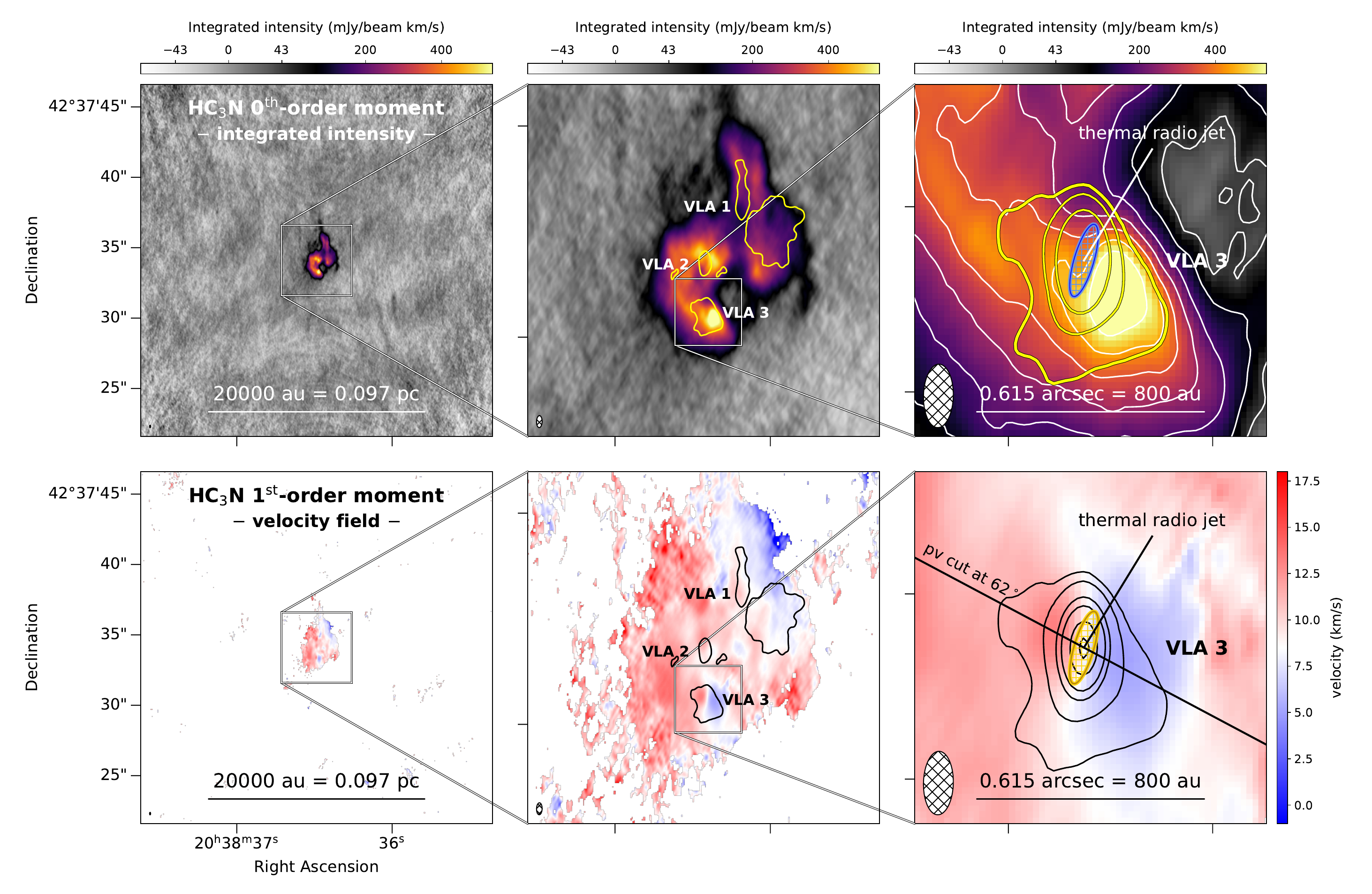}
\vspace{-0.5cm}
\caption{HC$_3$N\,(24--23) moment maps. Rows from top to bottom show: integrated intensity (0$^\mathrm{th}$ order); and velocity field (1$^\mathrm{st}$ order). The velocity interval used to generate the moment maps ranges from $-1.92$ up to $+18.18$~km~s$^{-1}$ (see Fig.~\ref{fig:HC3N_channels}). Columns from left to right show different zoomed-in views of the W75N(B) region, as described in Fig.~\ref{fig:continuum}. The ALMA 1.3~mm continuum emission (see Fig.~\ref{fig:continuum}) is shown in yellow contours in the top-row panels, and in black contours in the bottom-row panels. White contours in top-right panel shows the HC$_3$N integrated emission at levels 5, 10, 20, 30, 40, 50, and 60 times 9~mJy~beam$^{-1}$~km~s$^{-1}$, with the beam depicted by hatched ellipses. The blue and yellow ellipse in the right panels depicts the size and orientation of the thermal radio jet reported by \citet{CarrascoGonzalez2010a}. The diagonal solid black line in the right panel of second row shows the direction of the position-velocity (PV) cut, with PA$=62^\circ$, shown in Fig.~\ref{fig:PVcut}.}
\label{fig:HC3N_moments}
\end{figure*}

\begin{figure*}
\centering
\includegraphics[width=1.0\textwidth]{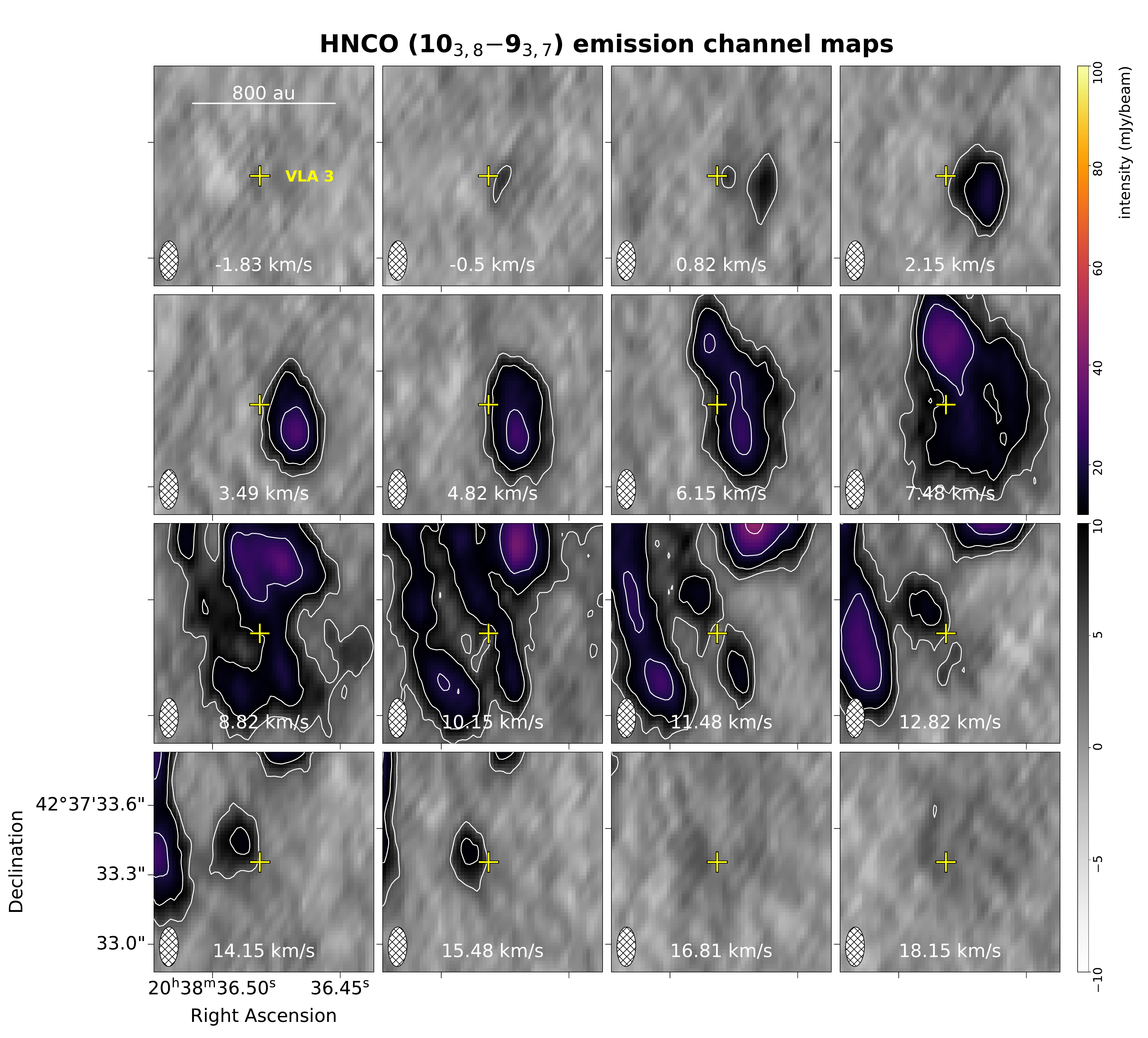}
\vspace{-0.5cm}
\caption{ALMA HNCO\,($10_{3,8}$--$9_{3,7}$) emission channel maps. From top-left to bottom-right the velocity of each channel, indicated at the bottom of each panel, increases from $-1.83$~km~s$^{-1}$ to $+18.15$~km~s$^{-1}$, in steps of $\approx1.35$~km~s$^{-1}$. The solid contours mark the HNCO emission at the levels 4, 8, 16, 32, and 64 times the rms of the image ($\mathrm{rms}=1.25$~mJy~beam$^{-1}$ in channels of $\approx1.35$~km~s$^{-1}$ width; beam $=0\farcs171\times0\farcs081$, PA$=-1\fdg4$, depicted with a hatched ellipse in each panel). The yellow cross marks the position of VLA3. The HNCO emission shifts from southwest with respect to VLA3 at blue-shifted velocities to northeast at red-shifted velocities (with the systemic velocity at V$_\mathrm{LSR}\simeq$~8.9~km~s$^{-1}$, see Sect.~\ref{sec:discussion}).}
\label{fig:HNCO_channels}
\end{figure*}

\begin{figure*}
\centering
\includegraphics[width=1.0\textwidth]{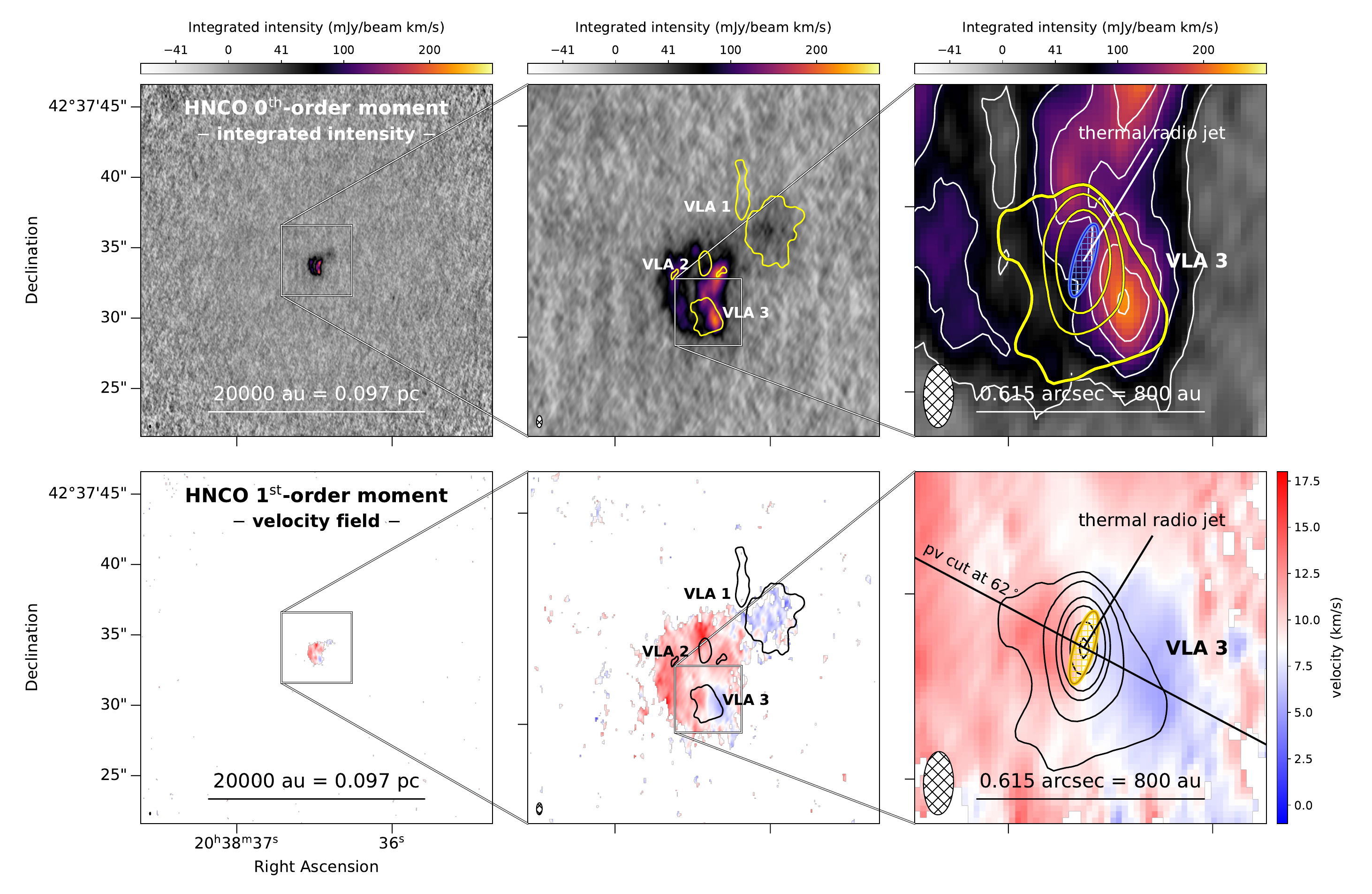}
\vspace{-0.5cm}
\caption{HNCO\,($10_{3,8}$--$9_{3,7}$) moment maps. Rows from top to bottom show: integrated intensity (0$^\mathrm{th}$ order); and velocity field (1$^\mathrm{st}$ order). The velocity interval used to generate the moment maps ranges from $-1.83$ up to $+18.15$~km~s$^{-1}$ (see Fig.~\ref{fig:HNCO_channels}). Columns from left to right show different zoomed-in views of the W75N(B) region, as described in Fig.~\ref{fig:continuum}. The ALMA 1.3~mm continuum emission (see Fig.~\ref{fig:continuum}) is shown in yellow contours in the top-row panels, and in black contours in the bottom-row panels. White contours in top-right panel shows the HNCO integrated emission at levels 5, 10, 15, 20, and 25 times 8.5~mJy~beam$^{-1}$~km~s$^{-1}$, with the beam depicted by hatched ellipses. The blue and yellow ellipse in the left panels depicts the size and orientation of the thermal radio jet reported by \citet{CarrascoGonzalez2010a}. The diagonal solid black line in the right panel of second row shows the direction of the position-velocity (PV) cut, with PA$=62^\circ$, shown in Fig.~\ref{fig:PVcut}.}
\label{fig:HNCO_moments}
\end{figure*}

\begin{figure*}
\centering
\includegraphics[width=1.0\textwidth]{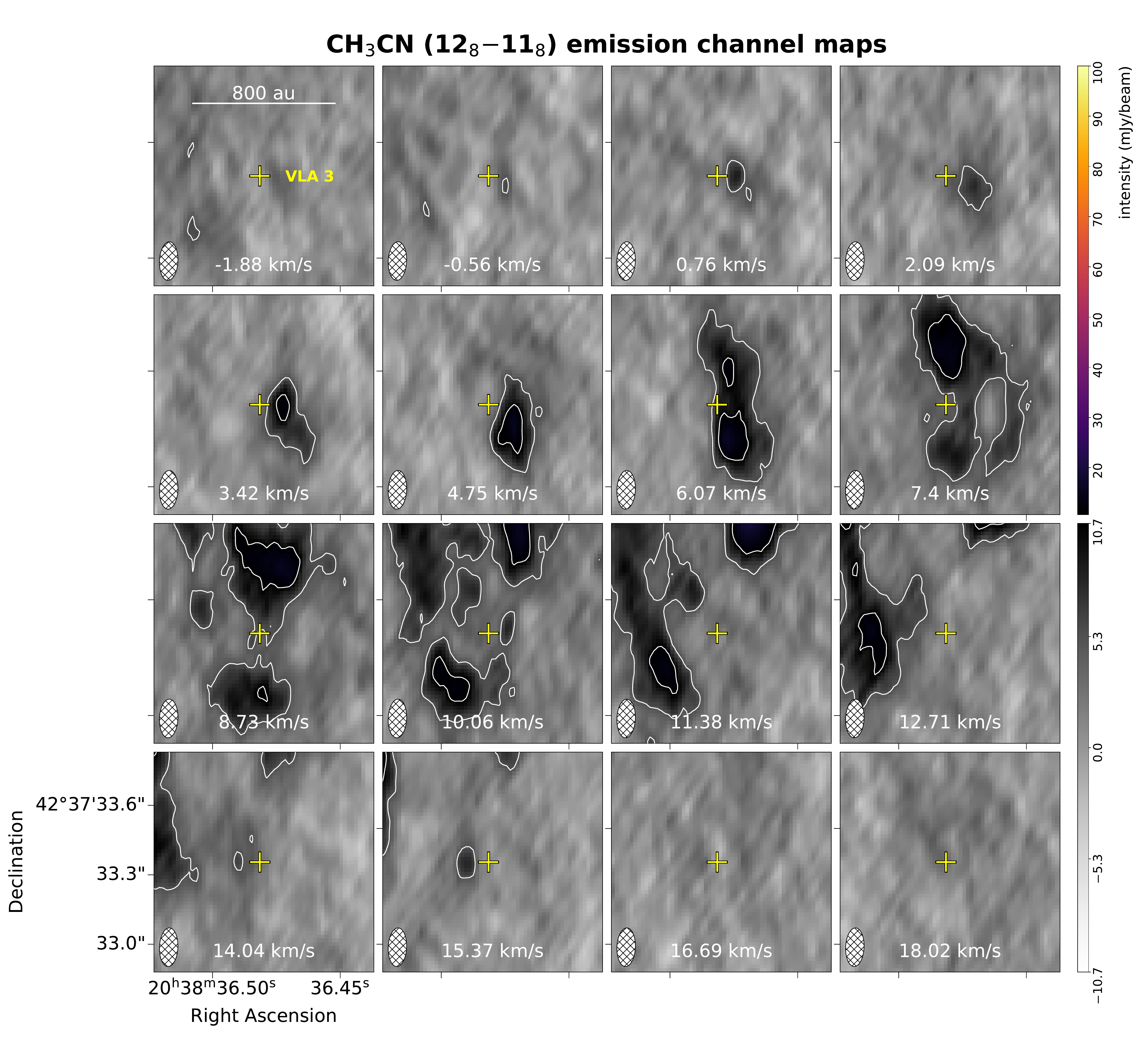}
\vspace{-0.5cm}
\caption{ALMA CH$_3$CN\,($12_{8}$--$11_{8}$) emission channel maps. From top-left to bottom-right the velocity of each channel, indicated at the bottom of each panel, increases from $-1.88$~km~s$^{-1}$ to $+18.02$~km~s$^{-1}$, in steps of $\approx1.35$~km~s$^{-1}$. The solid contours mark the CH$_3$CN emission at the levels 4, 8, 16, 32, and 64 times the rms of the image ($\mathrm{rms}=1.33$~mJy~beam$^{-1}$ in channels of $\approx1.35$~km~s$^{-1}$ width; beam $=0\farcs171\times0\farcs081$, PA$=-1\fdg4$, depicted with a hatched ellipse in each panel). The yellow cross marks the position of VLA3. The CH$_3$CN emission shifts from southwest with respect to VLA3 at blue-shifted velocities to northeast at red-shifted velocities (with the systemic velocity at V$_\mathrm{LSR}\simeq$~8.9~km~s$^{-1}$, see Sect.~\ref{sec:discussion}).}
\label{fig:CH3CN_channels}
\end{figure*}

\begin{figure*}
\centering
\includegraphics[width=1.0\textwidth]{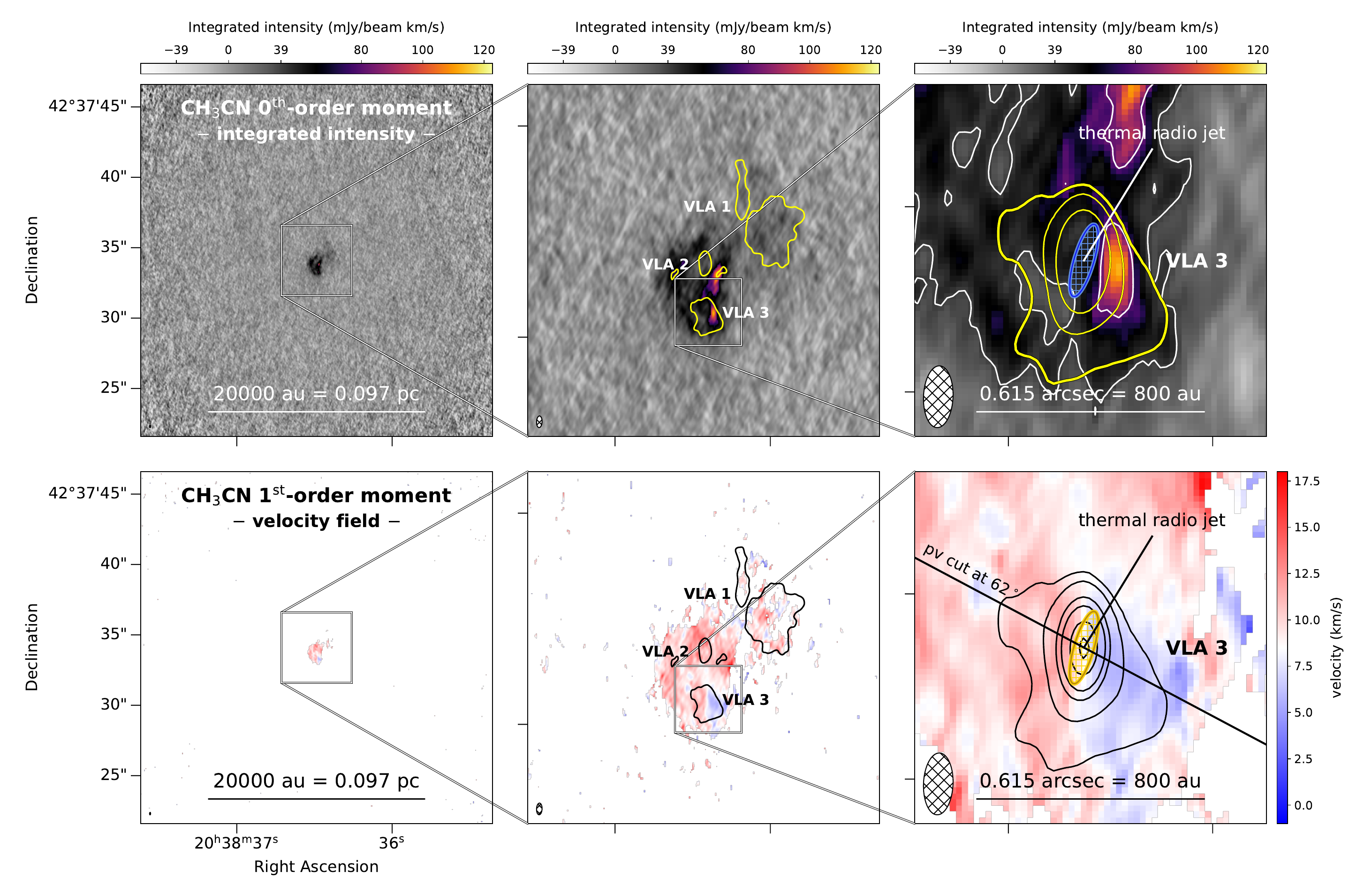}
\vspace{-0.5cm}
\caption{CH$_3$CN\,($12_{8}$--$11_{8}$) moment maps. Rows from top to bottom show: integrated intensity (0$^\mathrm{th}$ order); and velocity field (1$^\mathrm{st}$ order). The velocity interval used to generate the moment maps ranges from $-1.88$ up to $+18.02$~km~s$^{-1}$ (see Fig.~\ref{fig:CH3CN_channels}). Columns from left to right show different zoomed-in views of the W75N(B) region, as described in Fig.~\ref{fig:continuum}. The ALMA 1.3~mm continuum emission (see Fig.~\ref{fig:continuum}) is shown in yellow contours in the top-row panels, and in black contours in the bottom-row panels. White contours in top-right panel shows the CH$_3$CN integrated emission at levels 5 and 10 times 8.~mJy~beam$^{-1}$~km~s$^{-1}$, with the beam depicted by hatched ellipses. The blue and yellow ellipse in the right panels depicts the size and orientation of the thermal radio jet reported by \citet{CarrascoGonzalez2010a}. The diagonal solid black line in the right panel of second row shows the direction of the position-velocity (PV) cut, with PA$=62^\circ$, shown in Fig.~\ref{fig:PVcut}.}
\label{fig:CH3CN_moments}
\end{figure*}

Figures~\ref{fig:SiO_channels} to \ref{fig:CH3CN_moments} show the channel maps and the $0^\mathrm{th}$ and $1^\mathrm{st}$-order moment maps of the molecular species studied in this work (i.e., SiO, H$_2$CO, HC$_3$N, HNCO and CH$_3$CN). These figures complement similar figures shown in Figs.~\ref{fig:SO2_channels} and \ref{fig:SO2_moments} for the SO$_2$ species.

\begin{figure*}
\centering
\includegraphics[width=1.0\textwidth]{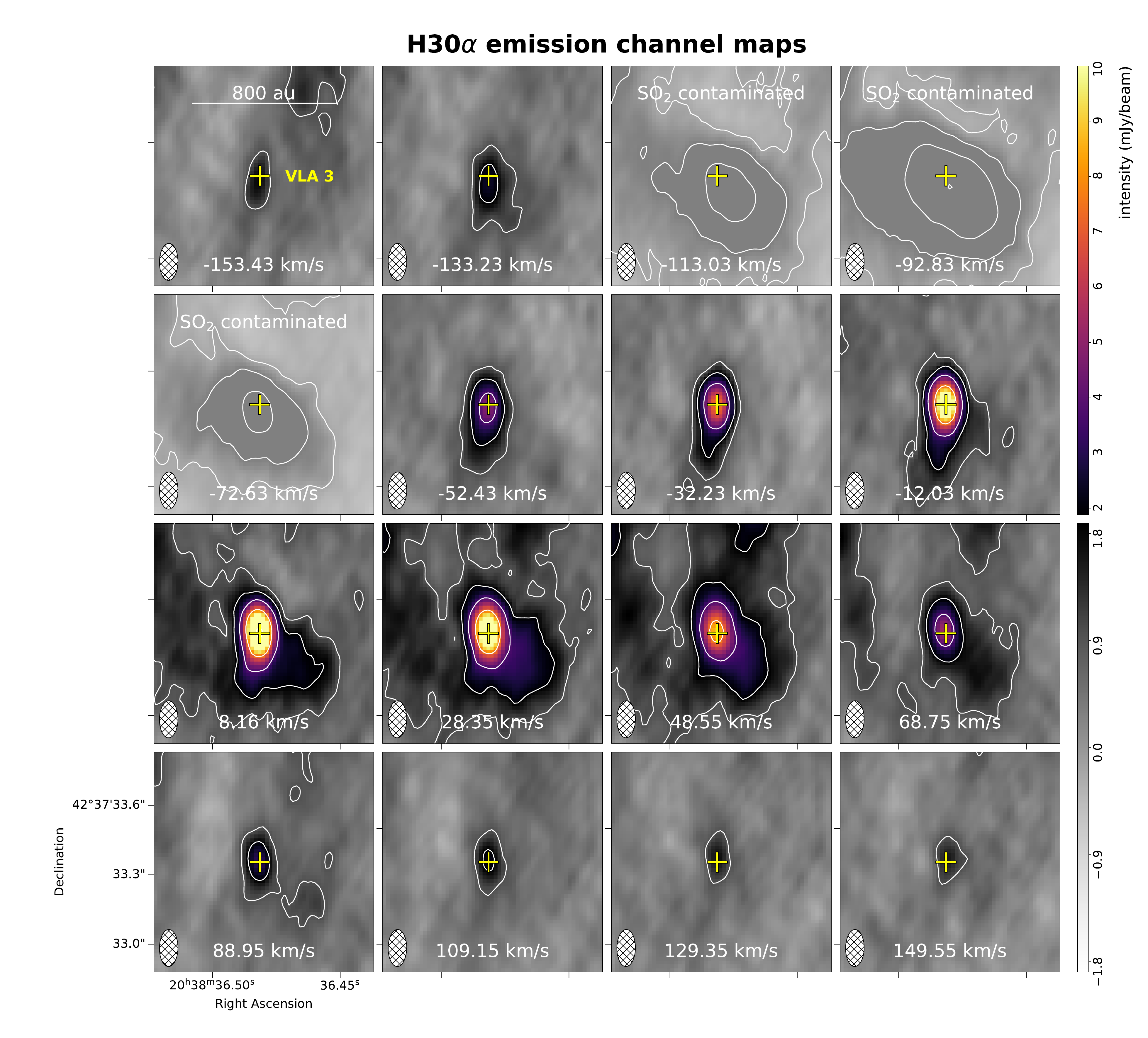}
\vspace{-0.5cm}
\caption{ALMA H30$\alpha$ emission channel maps. From top-left to bottom-right the velocity of each channel, indicated at the bottom of each panel, increases from $-153.43$~km~s$^{-1}$ to $+149.55$~km~s$^{-1}$, in steps of $\approx20$~km~s$^{-1}$. The solid contours mark the H30$\alpha$ emission at the levels 4, 8, 16, and 32, times the rms of the image ($\mathrm{rms}=0.24$~mJy~beam$^{-1}$ in channels of $\approx20$~km~s$^{-1}$ width; beam $=0\farcs160\times0\farcs080$, PA$=-0\fdg7$, depicted with a hatched ellipse in each panel). The yellow cross marks the position of VLA3. The emission in the channels from $-113$ to $-72$~km~s$^{-1}$ is dominated by the SO$_2$\,(14$_{3,11}$--14$_{2,12}$ ; v$_2$=1) transition at 231.980~GHz. No spatial shift in the peak emission is observed when moving from blue- to red-shifted velocities (with the systemic velocity at V$_\mathrm{LSR}\simeq$~8.9~km~s$^{-1}$, see Sect.~\ref{sec:discussion}).}
\label{fig:H30a_channels}
\end{figure*}




\bsp	
\label{lastpage}
\end{document}